\documentclass[preprint]{aastex}
\usepackage{emulateapj5}

\slugcomment{Accepted for publication in ApJ}

\shorttitle{VLBI images and cross-entropy optimizer}
\shortauthors{Caproni et al.}

\begin{document}


\title{Modeling very long baseline interferometric images with the cross-entropy global optimization technique}


\author{A. Caproni}
\affil{N\'ucleo de Astrof\'\i sica Te\'orica, Universidade Cruzeiro do Sul, R. Galv\~ao Bueno 868, Liberdade, 01506-000, S\~ao Paulo, SP, Brazil}
\email{anderson.caproni@cruzeirodosul.edu.br}

\author{H. Monteiro}
\affil{UNIFEI, Instituto de Ci\^encias Exatas, Universidade Federal
  de Itajub\'a, Av. BPS 1303, Pinheirinho, 37500-903, Itajub\'a, MG,
  Brazil}

\author{Z. Abraham}
\affil{Instituto de Astronomia, Geof\'\i sica e Ci\^encias Atmosf\'ericas, Universidade de S\~ao Paulo, R. do Mat\~ao 1226, Cidade Universit\'aria,\\
  05508-900, S\~ao Paulo, SP, Brazil}

\author{D. M. Teixeira}
\affil{Instituto de Astronomia, Geof\'\i sica e Ci\^encias Atmosf\'ericas, Universidade de S\~ao Paulo, R. do Mat\~ao 1226, Cidade Universit\'aria,\\
  05508-900, S\~ao Paulo, SP, Brazil}

\and

\author{R. T. Toffoli}
\affil{N\'ucleo de Astrof\'\i sica Te\'orica, Universidade Cruzeiro do Sul, R. Galv\~ao Bueno 868, Liberdade, 01506-000, S\~ao Paulo, SP, Brazil}




\begin{abstract}
We present a new technique for obtaining model fittings to very long baseline interferometric images of astrophysical jets. The method minimizes a performance function proportional to the sum of the squared difference between the model and observed images. The model image is constructed by summing $N_\mathrm{s}$ elliptical Gaussian sources characterized by six parameters: two-dimensional peak position, peak intensity, eccentricity, amplitude and orientation angle of the major axis. We present results for the fitting of two main benchmark jets: the first, constructed from three individual Gaussian sources, the second formed by five Gaussian sources. Both jets were analyzed by our cross-entropy technique in finite and infinite signal-to-noise regimes, the background noise chosen to mimic that found in interferometric radio maps. Those images were constructed to simulate most of the conditions encountered in interferometric images of active galactic nuclei. We show that the cross-entropy technique is capable of recovering the parameters of the sources with a similar accuracy to that obtained from the very traditional Astronomical Image Processing System Package task IMFIT when the image is relatively simple (e.g., few components). For more complex interferometric maps, our method displays superior performance in recovering the parameters of the jet components. Our methodology is also able to show quantitatively the number of individual components present in an image. An additional application of the cross-entropy technique to a real image of a BL Lac object is shown and discussed. Our results indicate that our cross-entropy model-fitting technique must be used in situations involving the analysis of complex emission regions having more than three sources, even though it is substantially slower than current model fitting tasks (at least 10,000 times slower for a single processor, depending on the number of sources to be optimized). As in the case of any model fitting performed in the image plane, caution is required in analyzing images constructed from a poorly sampled $(u,v)$ plane. 
\end{abstract}





\section{Introduction}

Very long baseline interferometry (VLBI) is to date the only tool available for the direct study of the structure and evolution of active galactic nuclei (AGNs) on parsec scales. While early observations were restricted to a few radio telescopes and required model fitting of the visibility function to obtain the source structure \citep{coh71}, good coverage of the $(u,v)$ plane of modern arrays and the development of efficient imaging techniques guarantee excellent maps of these sources \citep{rog74,red78,corn83,per84,she94}. 

Data with sparse $(u,v)$ plane coverage or calibration problems can produce images inappropriate for the study of quantitative aspects of AGNs, such as positions and proper motions of jet features (e.g., \citealt{pear99}). In this case, the analysis should be made by fitting a discrete number of model components (generally two-dimensional Gaussian) to the visibility data (e.g., \citealt{car93,kov05}), frequently corrected with self-calibrated techniques in the imaging process (\citealt{per95}). Fits in the $(u,v)$ plane have some advantages, such as error recognition, lack of spurious CLEAN artifacts, and full angular resolution capability (e.g., \citealt{pear99,lis01}). However, $(u,v)$ components do not necessarily correspond to real features in the image, but may simply be mathematical artifacts needed to reproduce complex brightness visibility data (e.g., \citealt{kel04,lis09b}). Besides, errors detected in the complex visibility data will not always produce measurable effects on the image plane \citep{eker99}, which means that analyses performed on the image plane are not necessarily less reliable than those carried out on the $(u,v)$ plane. Despite the intrinsic advantages and disadvantages, many works have used the image plane to extract information concerning the structural parameters of the observed sources (e.g., \citealt{kel04, wro04, rey09}).

Regardless of the plane where the analysis is taking place, the model fitting parameters are usually estimated from some maximum likelihood estimator, which  finds the best set of model parameters that minimize the residual differences between synthetic and real data. The convergence of such algorithms usually depends strongly on the initial estimation of the parameters and they are prone to finding non-global minimum solutions, especially if the object to be modeled is complex. In addition, the model-fitting routines generally used in interferometric astronomy (e.g., IMFIT, JMFIT, etc.) suffer from a limitation concerning the number of components to be fitted simultaneously in the image.

To overcome those limitations, we introduce a new and powerful technique: the cross-entropy method for global continuous optimization (hereafter CE). The method uses the image as input data, and searches for the optimal model parameters, selecting the best candidates among all solutions generated in each iteration and constructing new ones from them. There is no limitation to the number of Gaussian model-fitting components, and the algorithm is able to place self-consistently a practical limit on the number of model-fitting components, as will be discussed later in this work.

CE analysis was originally used in the optimization of complex computer simulation models involving rare events simulations \citep{rubi97}, and was modified by \citet{rubi99} to deal with continuous multi-extremal and discrete combinatorial optimization problems. Its theoretical asymptotic convergence has been demonstrated by \citet{marg04}, while \citet{kro06} studied its efficiency in solving continuous multi-extremal optimization problems. \citet{cap09} successfully applied the CE technique to determine precession model parameters of relativistic jets, while \citet{mon10} studied Galactic open clusters from CE optimization in color-magnitude diagrams. Other examples of application of the CE method are provided in \citet{deb05}.

The basic procedures involved CE optimization can be summarized as follows (e.g., \citealt{kro06}):

\begin{enumerate}
\item Random generation of the initial parameter sample, obeying pre-defined criteria;
\item Selection of the best samples based on some mathematical criterion;
\item Random generation of updated parameter samples from the previous best candidates to be evaluated in the next iteration;
\item Optimization process repeats steps (2) and (3) until a pre-specified stopping criterion is fulfilled.
\end{enumerate}

In this work, we validate our CE model-fitting algorithm from a variety of
benchmark tests built from synthetic jet components and apply it to a real image. The paper is structured as follows: in Section 2, we introduce the CE algorithm and its application to the problem of modeling interferometric radio images of astrophysical jets. The validation tests and their respective optimization results are discussed in Section 3. The application of our technique to a real image is presented in Section 4. Conclusions are presented in Section 5.

\section{The cross-entropy method for continuous optimization}

In this section, we introduce the cross-entropy method, describing how
to estimate the model-fitting parameters of jet knots present in the radio interferometric maps of astrophysical sources.

\subsection{Global View of the Optimization Process}

Let us consider an interferometric image composed of $N_x\times N_y$ pixels, where $N_x$ and $N_y$ are, respectively, the number of pixels in right ascension and declination coordinates. Assuming that such data can be described by an analytical model with $N_\mathrm{p}$ parameters $p_1, p_2, ..., p_{N_\mathrm{p}}$, we can use the CE optimization method to find the set of parameters ${\bf x}^*=(p^*_1,p^*_2,...,p^*_{Np})$ for which the model provides the best description of the data \citep{rubi99,kro06}. 

The optimization process is performed by randomly building $N$ independent sets of model parameters ${\bf X}=({\bf x}_1,{\bf x}_2,...,{\bf x}_N)$, where ${\bf x}_i=(p_{1i},p_{2i},...,p_{N{\mathrm{p}i}})$, and minimizing a performance function $S({\bf x})$ used to check the quality of the fit during the run process. In an ideal situation in which all parameters have converged to the exact solution, we must obtain $S({\bf x}^*)\rightarrow 0$.

To find the best solution from CE optimization, we start
by defining the parameter range in which the algorithm will search for
the best candidates: $p^\mathrm{min}_j\leq p_j(k) \leq p^\mathrm{max}_j$,
where $k$ represents the iteration number. Introducing
$\bar{p}_j(0)=(p^\mathrm{min}_j+p^\mathrm{max}_j)/2$ and
$\sigma_j(0)=(p^\mathrm{max}_j-p^\mathrm{min}_j)/2$, we can compute ${\bf
  X}(0)$ from:

\begin{equation}
  X_{ij}(0)=\bar{p}_j(0)+\sigma_j(0) G_{ij},
\end{equation}
where $G_{ij}$ is an $N\times N_\mathrm{p}$ matrix with random numbers
generated from a zero-mean normal distribution with standard deviation
of unity.

The next step is to calculate $S_i(0)$ for each set of ${\bf x}_i(0)$,
ordering them according to increasing values of $S_i$. Then the first
$N_\mathrm{elite}$ set of parameters is selected, i.e. the
$N_\mathrm{elite}$-sample with the lowest $S$-values, which will be labeled
as the elite sample matrix ${\bf X}^\mathrm{elite}(0)$.

We then determine the mean and standard deviation of the elite sample,
$\bar{p}^\mathrm{elite}_j(0)$ and ${\bf \sigma}^\mathrm{elite}_j(0)$
respectively, as:

\begin{equation}
  \bar{p}^\mathrm{elite}_j(0)=\frac{1}{N_\mathrm{elite}}\sum\limits_{i=1}^{N_\mathrm{elite}}X^\mathrm{elite}_{ij}(0),
\end{equation}

\begin{equation}
  {\bf \sigma}^\mathrm{elite}_j(0)=\sqrt{\frac{1}{\left(N_\mathrm{elite}-1\right)}\sum\limits_{i=1}^{N_\mathrm{elite}}\left[X^\mathrm{elite}_{ij}(0)-\bar{p}^\mathrm{elite}_j(0)\right]^2}.
\end{equation}

The matrix ${\bf X}$ at the next iteration is determined as:

\begin{equation}
  X_{ij}(1)=\bar{p}^\mathrm{elite}_j(0)+{\bf \sigma}^\mathrm{elite}_j(0) G_{ij},
\end{equation}

This process is repeated from Equation (2), with $G_{ij}$ regenerated
at each iteration. The optimization stops when the maximum number of iterations $k_\mathrm{max}$ is reached.

In order to prevent convergence to a sub-optimal solution due to the
intrinsic rapid convergence of the CE method, \citet{kro06} suggested
the implementation of a fixed smoothing scheme for
$\bar{p}^\mathrm{elite,s}_j(k)$ and ${\bf \sigma}^\mathrm{elite,s}_j(k)$:

\begin{equation}
  \bar{p}^\mathrm{elite,s}_j(k)=\alpha\bar{p}^\mathrm{elite}_j(k)+\left(1-\alpha\right)\bar{p}^\mathrm{elite}_j(k-1),
\end{equation}

\begin{equation}
  {\bf \sigma}^\mathrm{elite,s}_j(k)=\alpha_\mathrm{d}(k){\bf \sigma}^\mathrm{elite}_j(k)+\left[1-\alpha_\mathrm{d}(k)\right]{\bf \sigma}^\mathrm{elite}_j(k-1),
\end{equation}
where $\alpha$ is a smoothing constant parameter ($0<\alpha< 1$) and $\alpha_\mathrm{d}(k)$ is a dynamic smoothing parameter at the $k$th iteration:

\begin{equation}
  \alpha_\mathrm{d}(k)=\alpha-\alpha\left(1-k^{-1}\right)^q,
\end{equation}
where $q$ is an integer typically between 5 and 10 \citep{kro06}.

\subsection{Defining the Performance Function}

In the optimization problem we need to define a performance function based on the desired characteristics of the solution. Usually for continuous problems this is done by defining a likelihood function and then maximizing it, or requiring that the sum of the residuals squared be minimal.

In this work, we adopt a combination of the sum of the squared residuals and their respective variance. The motivation for this procedure is that there is no parametric form for determining the error for a given pixel of the interferometric radio images as is the case for typical photon-counting errors which are known to be Poissonian. Therefore we assumed that a good fit of jet sources would essentially produce small residuals, as well as a uniform spatial distribution of them, thus with minimal variance.

Let the quadratic residual $R_m(k)$ at a given pixel $m$ and iteration $k$ be defined as the squared difference between the observed image $I_m$ and the generated model image $M_m(k)$ at a $k$-iteration, i.e. $R_m(k)=\left[I_m-M_m(k)\right]^2$. The mean square residual value of the model fitting $\bar{R}(k)$ can be calculated from:

\begin{equation}
 \bar{R}(k)=\frac {1} {N_\mathrm{pixel}}\left[\sum\limits_{m=1}^{N_\mathrm{pixel}} R_m(k)\right].
\end{equation}

We choose to rank our tentative model images obtained from the $6N_\mathrm{s}$ parameters ${\bf x}_i(k)$ at iteration $k$ through the performance function:
 
\begin{equation}
  S_\mathrm{prod}({\bf x}_i,k)=\bar{R}(k)\times\frac {1} {N_\mathrm{pixel}}
  \left[\sum\limits_{m=1}^{N_\mathrm{pixel}} \left(R_m(k)-\bar{R}(k)\right)^2\right],
\end{equation}
in which the product between $\bar{R}$ and the variance-like factor was used. Note also that, if $I_m$ and $M_m$ are expressed in terms of Jy beam$^{-1}$, $S_\mathrm{prod}$ has units of Jy$^6$ beam$^{-6}$.

It is important to emphasize that we have tested other functional forms for the CE performance function but Equation (9) was more efficient in more complex problems. We believe that the difference in performance is essentially due to the fact that Equation (9) transmits directly any change in the mean and variance of the residuals in all iteration steps of the optimization process.

\subsection{Estimation of the model fitting parameters and their uncertainties}

As commented previously, CE optimization generates $N$ random tentative solutions at each iteration $k$, selecting the best $N_\mathrm{elite}$ set of model parameters in terms of the values of $S_\mathrm{prod}$.

In all optimization process presented in this work, we first chose the number of sources $N_\mathrm{s}$ to be model fitted on the image and applied CE optimization three times for each choice of $N_\mathrm{s}$. After finalizing the three optimization runs, we selected the optimization that best minimized $S_\mathrm{prod}$ in order to determine the best values of the $6N_\mathrm{s}$ Gaussian parameters $p_i^*$, as well as their respective uncertainties $\sigma_{p_i^*}$ ($i$ varying from 1 to $6N_\mathrm{s}$) as follows:

\begin{equation}
p_i^*=\sum\limits_{k=1}^{k_\mathrm{max}}w_kp_{ik}\left(\sum\limits_{k=1}^{k_\mathrm{max}}w_k\right)^{-1}
\end{equation}
and

\begin{equation}
\sigma_{p^*_{i}}^2=\sum\limits_{k=1}^{k_\mathrm{max}}w_k\left(p_{ik}-p^*_{i}\right)^2\left(\sum\limits_{k=1}^{k_\mathrm{max}}w_k\right)^{-1},
\end{equation}
where $k_\mathrm{max}$ is the maximum number of iterations used in the optimized model fitting, and $p_{ik}$ represents the set of model parameters that produce the minimum value of $S_{\mathrm{prod}_k}$ among all tentative solutions at iteration $k$ and $w_k=S^{-2}_{\mathrm{prod}_k}$. The power index -2 in the definition of $w_k$ was adopted in order to make the tentative solutions with the lowest values of $S_{\mathrm{prod}_k}$ more important in the calculation of $p_i^*$.

The same procedure was used to estimate the value of $S_\mathrm{prod}$ associated with $p_i^*$, except for $w_k=\sigma^{-4}_{S_{\mathrm{prod}_k}}$, where $\sigma^2_{S_{\mathrm{prod}_k}}$ is the variance of $S_{\mathrm{prod}_k}$ among all tentative solutions at an iteration $k$.

\section{Validating the CE model-fitting method}

To validate the fitting technique, we used synthetic images generated from a previously known set of parameters. Such images were composed of $N_\mathrm{s}$ two-dimensional elliptical Gaussian sources, characterized by the Gaussian center peak positions $(x,y)$, semi-major axis $a$, eccentricity $\epsilon$, position angle $\theta$, and peak intensity $I_0$. 

We created two synthetic jets, labeled J1 and J2, with a bright main central source (emulating an unresolved core as seen in real interferometric images) and additional components with a range of sizes and with different orientations, intensities, and degrees of superposition. We took them as being representative of typical jet sources found in the literature. 

We also ran simpler tests, such as single source, fully separated set of sources, etc., but we do not present them here because they do not give any more insight into the performance of the method. In all these simple cases the method performed extremely well, finding the correct solution with considerable ease.

Our method was implemented in such a way that the fitting was entirely automatic. This has opened the possibility of performing fits for a wide range and number of components, analyzing their quality a posteriori based on more quantifiable characteristics. In particular, we have used mainly the best value of the fitness function $S_\mathrm{prod}^*({\bf x}_i)$ and the characteristics of the residual maps to decide the ideal number of sources, as can be seen in the following sections. 

It is important to emphasize that we have adopted $k_\mathrm{max}=2000N_\mathrm{s}$, $\alpha=0.9$, $q=5$, $N=N_0(6N_\mathrm{s})^2$ (with $30\lesssim N_0\lesssim 50$) and $N_\mathrm{elite}=0.05N$\footnote{\citet{cap11} adopted $k_\mathrm{max}=5000$ and $N=50(6N_\mathrm{s})^2$.} in all validation tests presented in this work, in agreement with those suggested by \citet{kro06} and Caproni et al. (2009). Additional tests were made using different values for those CE parameters but they did not improve the overall performance of our technique. The CE optimizations presented in next sections were run on a 2.7 GHz processor, taking typically from a couple of minutes up to days to be fully completed, depending on the number of elliptical Gaussian components assumed in the model fitting.

\subsection{Noiseless Synthetic Benchmark Tests}

This section is devoted to showing the performance of our CE model fitting in ideal situations where there is no background noise present, i.e., the signal-to-noise ratio (S/N) tends to infinity.

\subsubsection{The Noiseless Synthetic Jets J1 and J2}

\begin{figure}
     \epsscale{0.25}
	  {\plotone{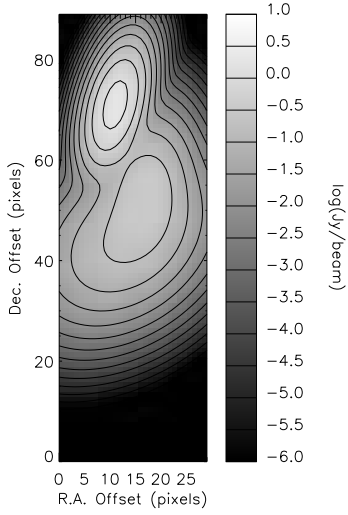}}
          \caption{Image of the synthetic jet built from three discrete elliptical Gaussian components (see Table \ref{tab_J1} for the jet component parameters). The gray-scale map as well as the contours is in logarithm scale in units of Jy beam$^{-1}$.}
      \label{im_3gauss_no_noise}
   \end{figure}

Jet J1 contains three elliptical Gaussian sources ($N_\mathrm{s}=3$) for which peak intensity/size decreases/increases with the core-component distance. Their respective parameters are shown in Table \ref{tab_J1}. The synthetic image of J1 with a size of 30 $\times$ 90 pixels in right ascension and declination, respectively, is displayed in Figure \ref{im_3gauss_no_noise}. While the brightest component (mimicking the core of the object) is relatively detached from the others, the remaining two components present a moderate degree of superposition, mainly because of the size of component 3.

\begin{figure}
     \epsscale{0.30}
	  {\plotone{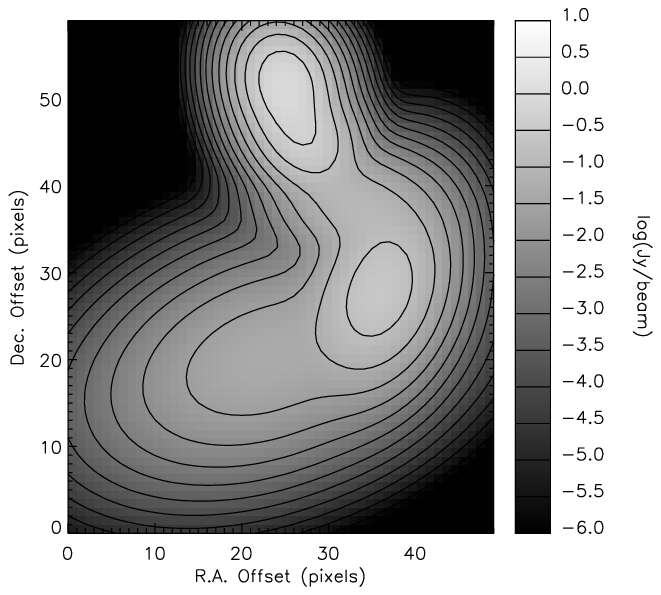}}
          \caption{Image of the synthetic jet built from five discrete elliptical Gaussian components (see Table \ref{tab_J2} for the jet component parameters). The gray-scale map as well as the contours is in logarithm scale in units of Jy beam$^{-1}$.}

      \label{im_5gauss_no_noise}
   \end{figure}

Jet J2 is composed of five distinct elliptical Gaussian sources ($N_\mathrm{s}=5$), with a small central source and the others at increasing distances, and larger sizes, as usually observed in astrophysical jets. The source parameters are listed in Table \ref{tab_J2}, while the resulting map covering an area of 50 $\times$ 60 pixels in right-ascension-declination plane is shown in Figure \ref{im_5gauss_no_noise}. The distances between the components were chosen to introduce different degrees of spatial superposition. We also put a weaker source between two bright ones (source number 3) with high degree of superposition to evaluate the ability of the method to separate sources with intensities slightly above the wing intensities of neighboring sources.

\subsubsection{The CE Model-fitting Results for J1 and J2 Noiseless Maps}

The CE model-fitting process generates elliptical Gaussian sources with peak positions within the observed image boundaries and from a large span of parameter space as follows: $2.5\leq a \mathrm{(pixel)}\leq 40$, $0\leq \epsilon\leq 0.95$ and $-90\degr\leq \theta \leq 90\degr$ (positive values for north to east direction) for both J1 and J2 jets. In relation to the peak intensity, the limits are $0.0002\leq I_0$ (Jy beam$^{-1}$)$\leq 2.1$ for J1 and $0.0001\leq I_0$ (Jy beam$^{-1}$)$\leq 1.2$ for J2.

The values of the model fitting parameters obtained from our CE global optimization technique for J1 and J2 benchmark tests are shown, respectively, in Tables \ref{tab_J1_modfit_no_noise} and \ref{tab_J2_modfit_no_noise}. Those values, as well as their respective uncertainties, were calculated from the procedures described in Section 2.3. The number of sources adopted in each optimization was varied from two to seven for J1 and three to seven for J2 (justification for this procedure is given in the next subsection).

A quick comparison between Tables \ref{tab_J2} and \ref{tab_J2_modfit_no_noise} reveals that our algorithm fits tentatively the most intense sources present in the image if $N_\mathrm{s}<5$. In other words, components 1 (core), 2 and 4 are always present in the modeling, even though their recovered parameters are not necessarily correct in all cases. If we employ $N_\mathrm{s}>5$, the algorithm places the extra Gaussian sources close to the location of the most intense sources. Except for the threshold value $N_\mathrm{s}=3$, the same considerations are valid for the comparison between Tables \ref{tab_J1} and \ref{tab_J1_modfit_no_noise} related to jet J1. 

Comparison between the values of Table \ref{tab_J2} and the five-source entries of Table \ref{tab_J2_modfit_no_noise} reveals an excellent agreement between them. The same is true in the case of jet J1 (compare the three-source entries of Tables \ref{tab_J1} and \ref{tab_J1_modfit_no_noise}), for which our algorithm recovered the structural parameters of the three sources with a tantalizing zero relative error.

\subsubsection{Estimating the Number of Jet Components from the CE Model-Fitting}

There is an important aspect that must be addressed when any model fitting technique is applied to an interferometric image: how many components are supposed to be used in the fit? Unfortunately the answer is that there is no ultimate rule to apply in order to determine the number of components to be fitted in the image (even a minimum or a maximum limit). Perhaps the most reasonable and conservative criterion could be the one based on the Occam's Razor principle, adopting the lowest number of sources for a reasonable fit. This strategy can be found in some previous works that deals with model fitting of VLBI images (e.g., \citealt{hom01}).

\begin{figure}
     \epsscale{0.25}
	  {\plotone{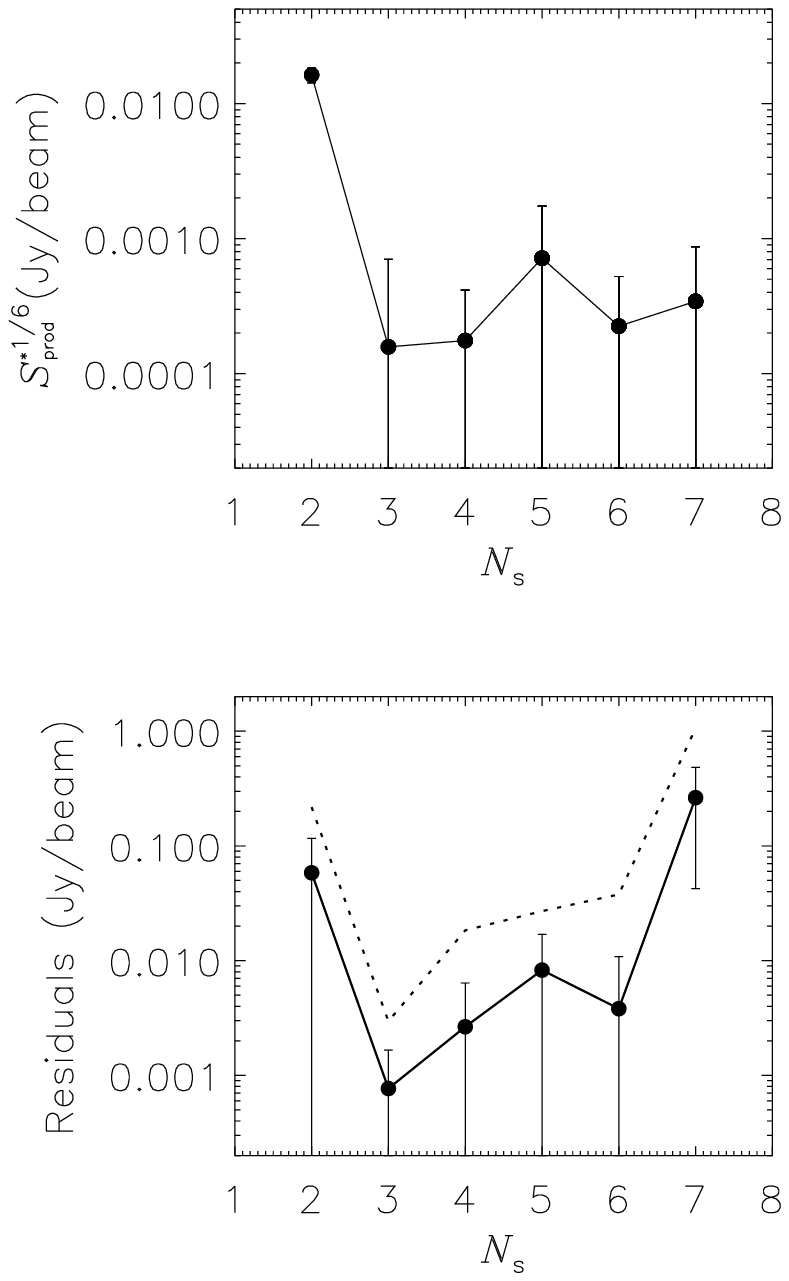}}
          \caption{Quantitative analyses of the model fittings in the case of the noiseless synthetic image of J1. {\it Top}: one-sixth of the performance function as a function of the number of sources used in the CE optimization. Taking into account the error bars, there is a plateau-like structure for $N_\mathrm{s}\ge 3$. {\it Bottom:} Behavior of the mean value of the residuals in terms of the number of sources used in the CE optimization. The error bars are the standard deviation of the corresponding residual images. The dotted line refers to the maximum value of the residuals. The mean and maximum values reach their minimum at $N_\mathrm{s}=3$.}
      \label{fitness_J1_no_noise}
   \end{figure}

\begin{figure}
     \epsscale{0.25}
	  {\plotone{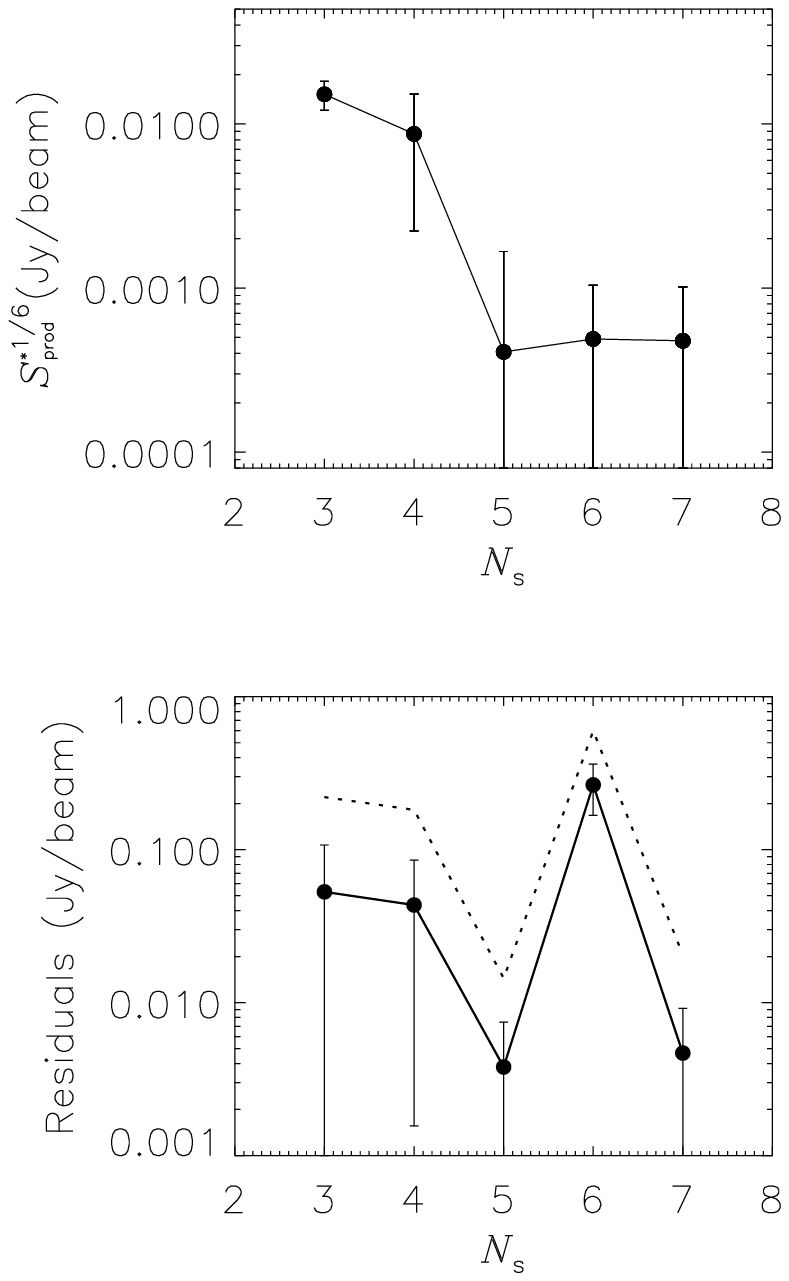}}
          \caption{Quantitative analyses of the model fittings in the case of the noiseless synthetic image of J2. {\it Top:} One-sixth of the performance function as a function of the number of the sources used in the CE optimization. Taking into account the error bars, there is a plateau-like structure for $N_\mathrm{s}\ge 5$. {\it Bottom:} Behavior of the mean value of the residuals in terms of the sources used in the CE optimization. The error bars correspond to the standard deviation of the residual images. The dotted line refers to the maximum value of the residuals.}
      \label{fitness_J2_no_noise}
   \end{figure}

To verify if our algorithm can determine quantitatively the number of sources, or at least the minimum number of components that should be assumed in the modeling, we have varied the number of Gaussian sources in the CE optimization procedure. We show in the top panels of Figures \ref{fitness_J1_no_noise} and \ref{fitness_J2_no_noise} the one-sixth values\footnote{As has been mentioned in Section 2.2, the performance function $S_\mathrm{prod}^*$ has units of Jy$^6$ beam$^{-6}$.} of the performance function $S_\mathrm{prod}^*$ as a function of the number of sources assumed, respectively, in the model fitting of J1 and J2. The error bars are large and only allow us to conclude that there is possibly a plateau-like structure after $N_\mathrm{s}\simeq 3$ for J1 and $N_\mathrm{s}\simeq 5$ for J2. This suggests a crude estimate for the minimum value for the number of sources present in the image. This is a slight advance in relation to previous works since our technique allowed us to put a lower limit on the number of the sources, based only on the values of our performance function.

With the aim of improving our estimate of $N_\mathrm{s}$, we plotted in the bottom panels of Figures \ref{fitness_J1_no_noise} and \ref{fitness_J2_no_noise} the mean  and the maximum values of the residual images obtained from our CE model fittings for J1 and J2 respectively. These quantities indicate that there is a minimum at  $N_\mathrm{s}=3$ for J1 and $N_\mathrm{s}=5$ and 7 for J2, which means we could not unequivocally constrain $N_\mathrm{s}$ for jet J2. However, looking carefully at Table \ref{tab_J2_modfit_no_noise}, we can see that the same Gaussian sources for $N_\mathrm{s}=5$ were also found in the model fitting employing $N_\mathrm{s}=7$. The extra two components are too dim, compatible with zero-intensity taking into account the uncertainties. Therefore,  the ambiguity in the correct number of the components is definitely eliminated.

It is important to emphasize that similar behavior was found for jet J1, the three sources for $N_\mathrm{s}=3$ having also been detected by our CE optimizations in the cases of $N_\mathrm{s}=4, 5$ and 6. Thus, we believe that the procedures mentioned above can be very useful and effective in putting a quantitative constraint on the number of sources present in any interferometric image.

\subsection{Noise Validation Test}

In the previous section, we checked the performance of our CE model-fitting technique in recovering the structural parameters of synthetic sources in the limit of infinite S/N. However, this idealized situation is definitely far from those found when dealing with real interferometric radio maps, in which the S/R has a finite value. Thus, it is necessary to verify if our technique also works in such conditions.

The question that arises is how to perform tests that include a realistic noise background. In the VLBI images obtained by the CLEAN method the presence of structured noise is evident, in other words, noise not originating in Gaussian or Poissonian distributions. Because there is no simple parametric expression for the noise in those images, we carried out the following procedure to obtain our finite-S/R benchmark images: 

\begin{enumerate}
  \item Random selection of a real interferometric image from the data archive of the MOJAVE program\footnote{\url{http://www.physics.purdue.edu/MOJAVE/index.html}};
  \item Selection of a region of this image where no apparent sources are present;
  \item Extraction of a noise image of a given size from that region;
  \item Construction of a synthetic image composed by $N_\mathrm{s}$ Gaussian sources, with the central one having the same characteristics as the CLEAN beam parameters found in the header of the noise image;
  \item Addition of the noise and synthetic images to form our new benchmark map.
\end{enumerate}

Note that this procedure simulates accurately the thermal noise characteristics of the image, as well as some of the CLEAN algorithm effects. The effects of wide gaps in the interferometric $(u,v)$ coverage, (e.g., aliasing, sinusoidal ripples, etc.) are not necessarily present in our benchmark images. However, a rigorous analysis of real CLEAN images can reveal the presence of such (usually weak) artifacts (e.g., \citealt{eker99, lis01}), allowing the adoption of some additional strategy to avoid misinterpretation of the data image. As regards rippling features, changes in the CLEAN algorithm can substantially reduce or even remove them from the final CLEAN image \citep{corn83}. Similarly, \citet{pfko98} proposed a different deconvolution method to translate raw complex visibility data into an image that eliminates aliasing effects.

\begin{figure*}
     \epsscale{0.5}

	  {\plotone{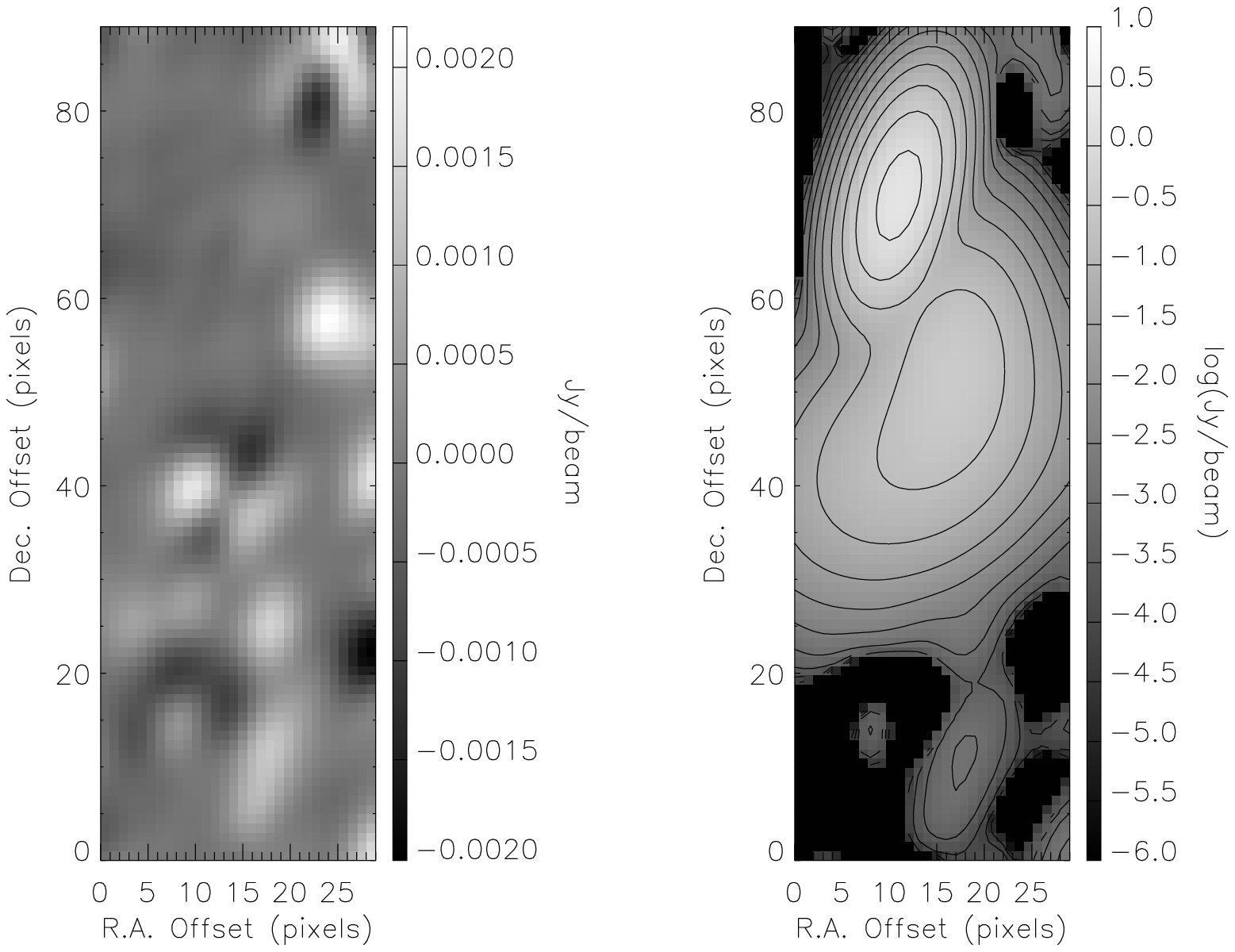}}
          \caption{{\it Left:} Noise pattern extracted from a MOJAVE project image used to build the background noise. RMS of this noise image is 0.515 mJy beam$^{-1}$. {\it Right:} Image of the noise validation test composed of the summation of the contribution of three elliptical Gaussian sources of jet J1 and the background noise shown in the left panel.}
      \label{noise_plus_map_J1}
   \end{figure*}

\begin{figure*}
     \epsscale{0.7}

	  {\plotone{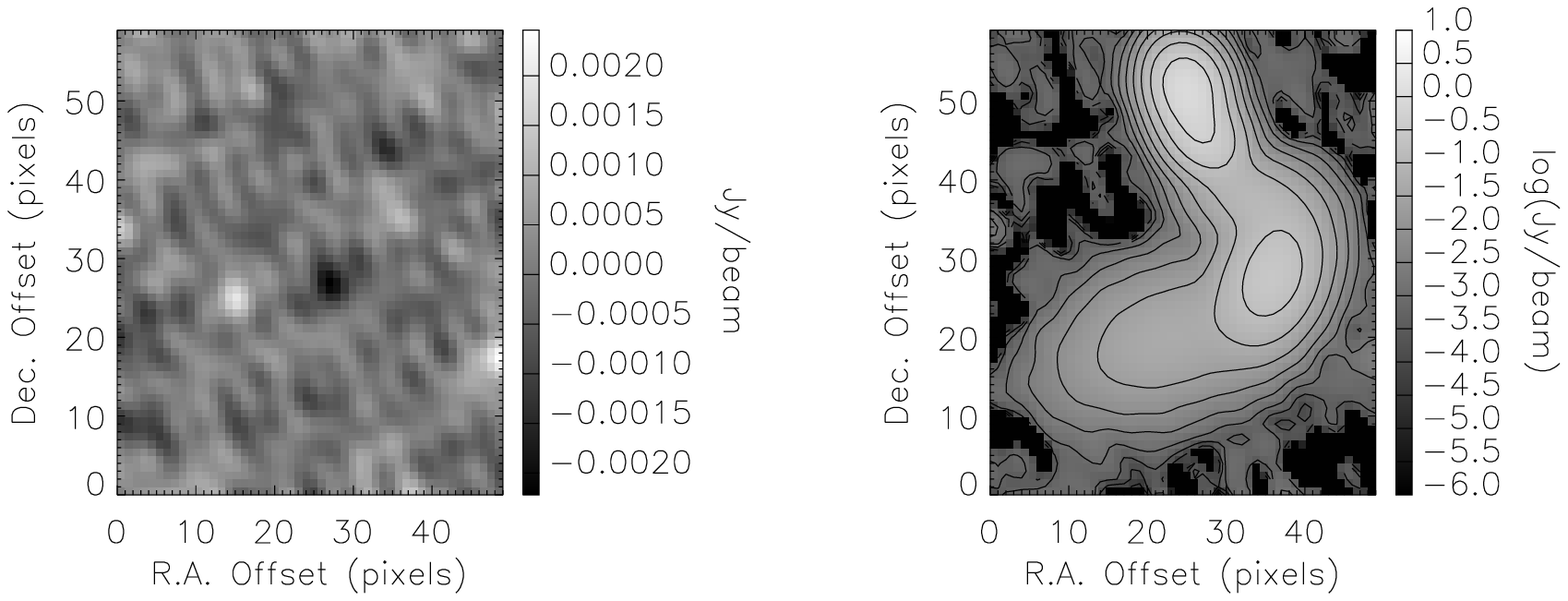}}
          \caption{{\it Left:} Noise pattern extracted from a MOJAVE project image used to build the background noise. RMS of this noise image is 0.463 mJy beam$^{-1}$. {\it Right:} Image of the noise validation test composed of the summation of the contribution of five elliptical Gaussian sources of jet J2 and the background noise shown in the left panel.}
      \label{noise_plus_map_J2}
   \end{figure*}

For the synthetic sources, we decided to use the same jet knots J1 and J2 as used in the previous section. The noise pattern, and the resulting new benchmark images are shown in Figures \ref{noise_plus_map_J1} and \ref{noise_plus_map_J2}. Although this procedure might not produce a synthetic image with precise representation of the expected noise pattern, we believe that it is the best way to generate a more realistic image to evaluate the influence of noise in the CE model-fitting optimization process.

In analogy with the previous section, our CE model-fitting algorithm examined the whole image in the $x-$ and $y-$ coordinates, searching for Gaussian sources with structural parameters in the same range as used in the noiseless cases. To guarantee an unbiased solution we ran the CE algorithm for different numbers of sources, varying from two up to seven. As in the noiseless case, we adopted $N_0=30$ in the optimization of the J1 image. For J2, we had to increase $N_0$ from 30 to 50 to guarantee a good performance of our algorithm.

Following Section 2.3, we obtained the optimal values for the model parameters and their respective uncertainties, which are displayed in Tables \ref{tab_J1_modfit_noise} and \ref{tab_J2_modfit_noise}. As in the noiseless cases, comparison between Tables \ref{tab_J1_modfit_noise} and \ref{tab_J2_modfit_noise} reveals that our algorithm fits tentatively the most intense sources present in the image when the number of sources assumed in the CE model fitting is smaller than the correct value. If the assumed number of the sources is larger than the right one, the algorithm puts the extra dim Gaussian sources close to the location of the most intense sources.

\begin{figure*}
     \epsscale{1}

	  {\plotone{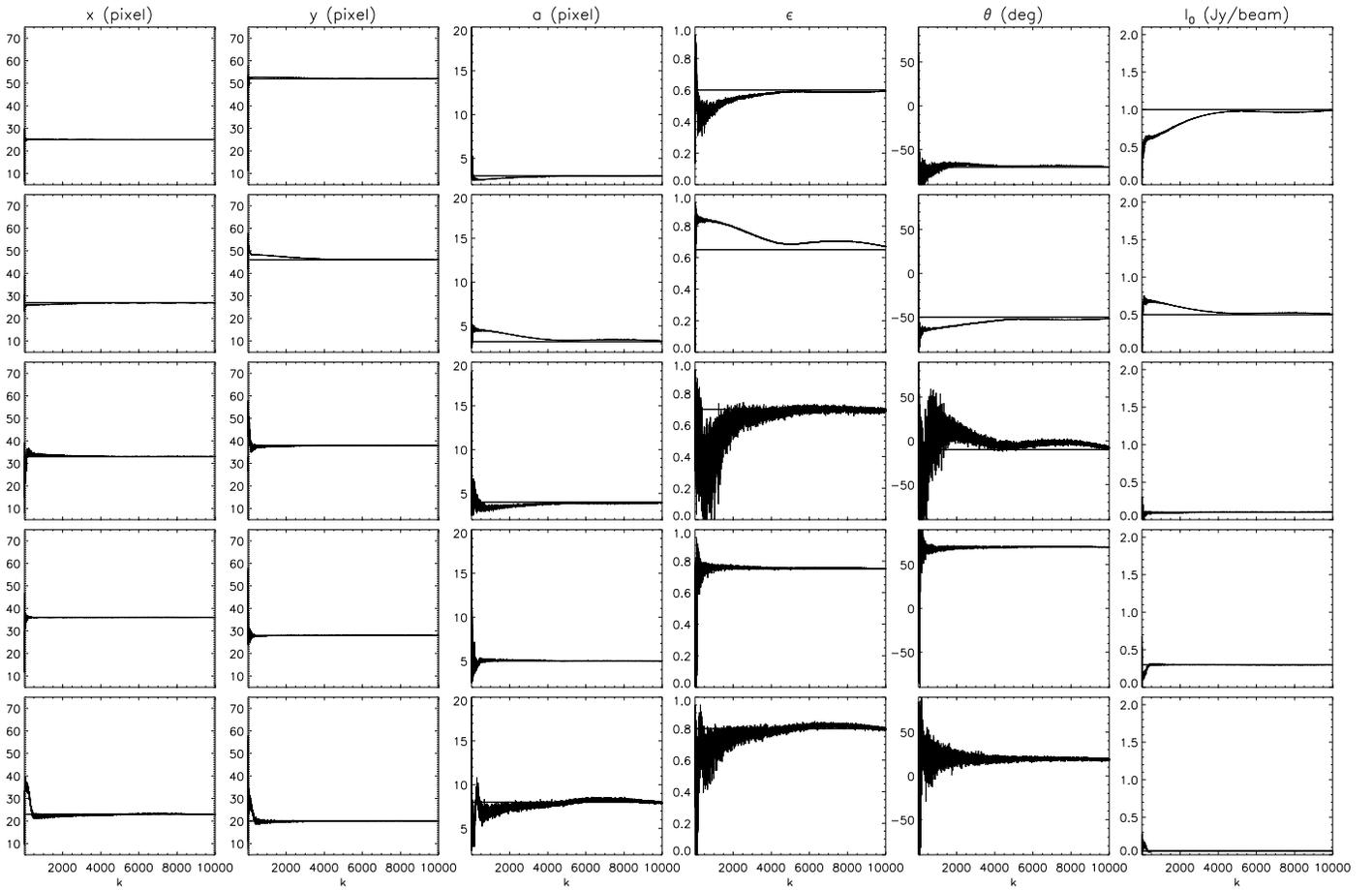}}
          \caption{Evolution of the model parameters of the Gaussian sources of the noisy jet J2 as a function of the iteration number $k$. The solid horizontal lines represent the expected values for that benchmark test (see Table \ref{tab_J2}).}
      \label{Param_converg_J2}
   \end{figure*}

As an example, we show in Figure \ref{Param_converg_J2} the evolution of the values of the CE-optimized model parameters of the noisy jet J2 as a function of the iteration number. The parameter values reach a steady-state-like behavior at iterations below $k_\mathrm{max}$ after converging to the expected values. Similar findings are also observed for the other benchmark tests discussed in this work.

\begin{figure*}
     \epsscale{0.90}

	  {\plotone{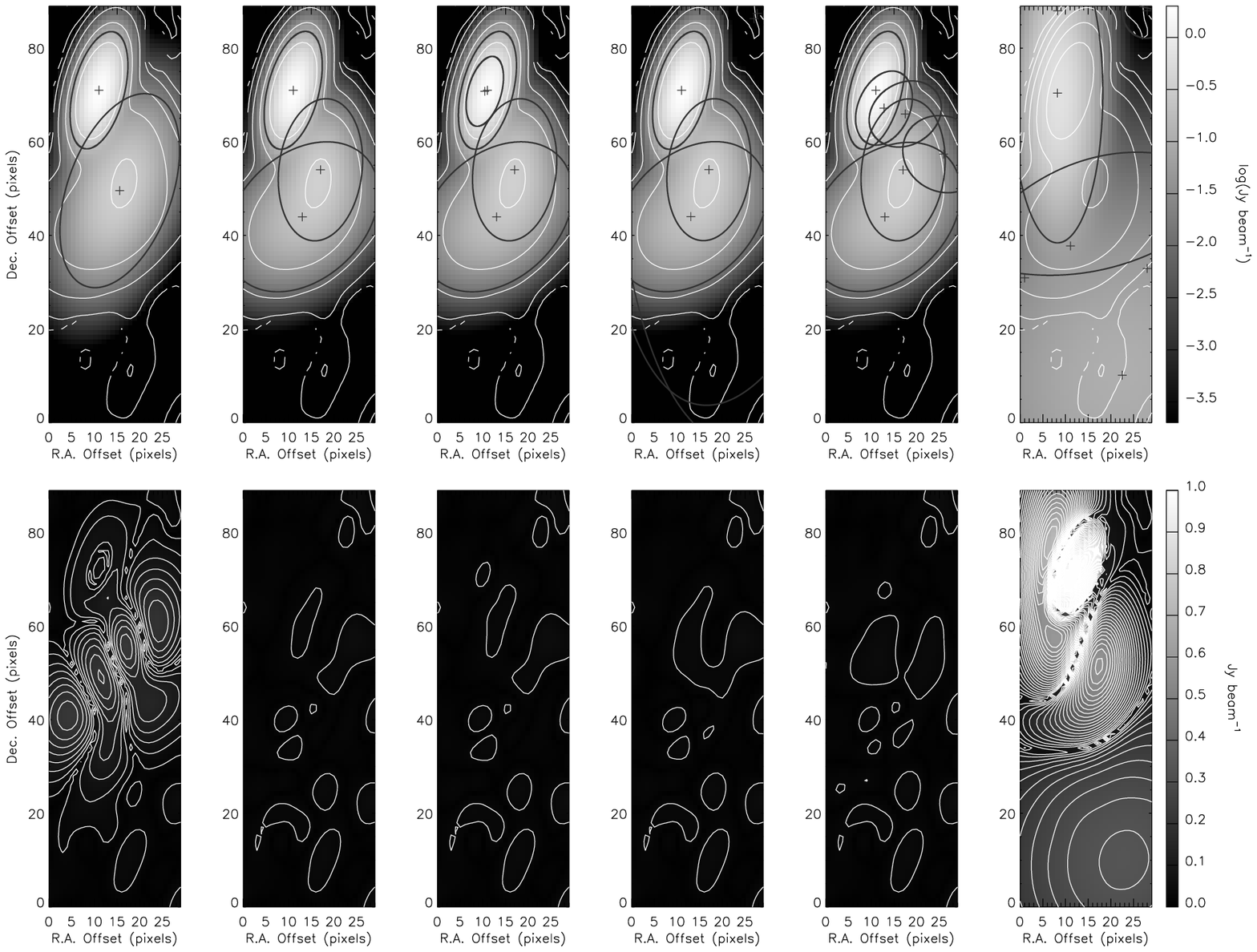}}
          \caption{Results of the model fitting for the synthetic image of the jet J1 with the addition of a realistic noise pattern obtained after varying the number of sources from two to seven (left-to-right direction). $\it Top:$ Contour lines are of the original image shown in Figure \ref{noise_plus_map_J1}, the gray-scale image is constructed from the fitted source parameters, and the dark ellipses are the contours of the individual fitted sources (respective centers marked with crosses) at the FWHM. $\it Bottom:$ Respective residual maps in linear scale.}
      \label{img_noise_synth_J1}
   \end{figure*}

\begin{figure*}
     \epsscale{0.9}

	  {\plotone{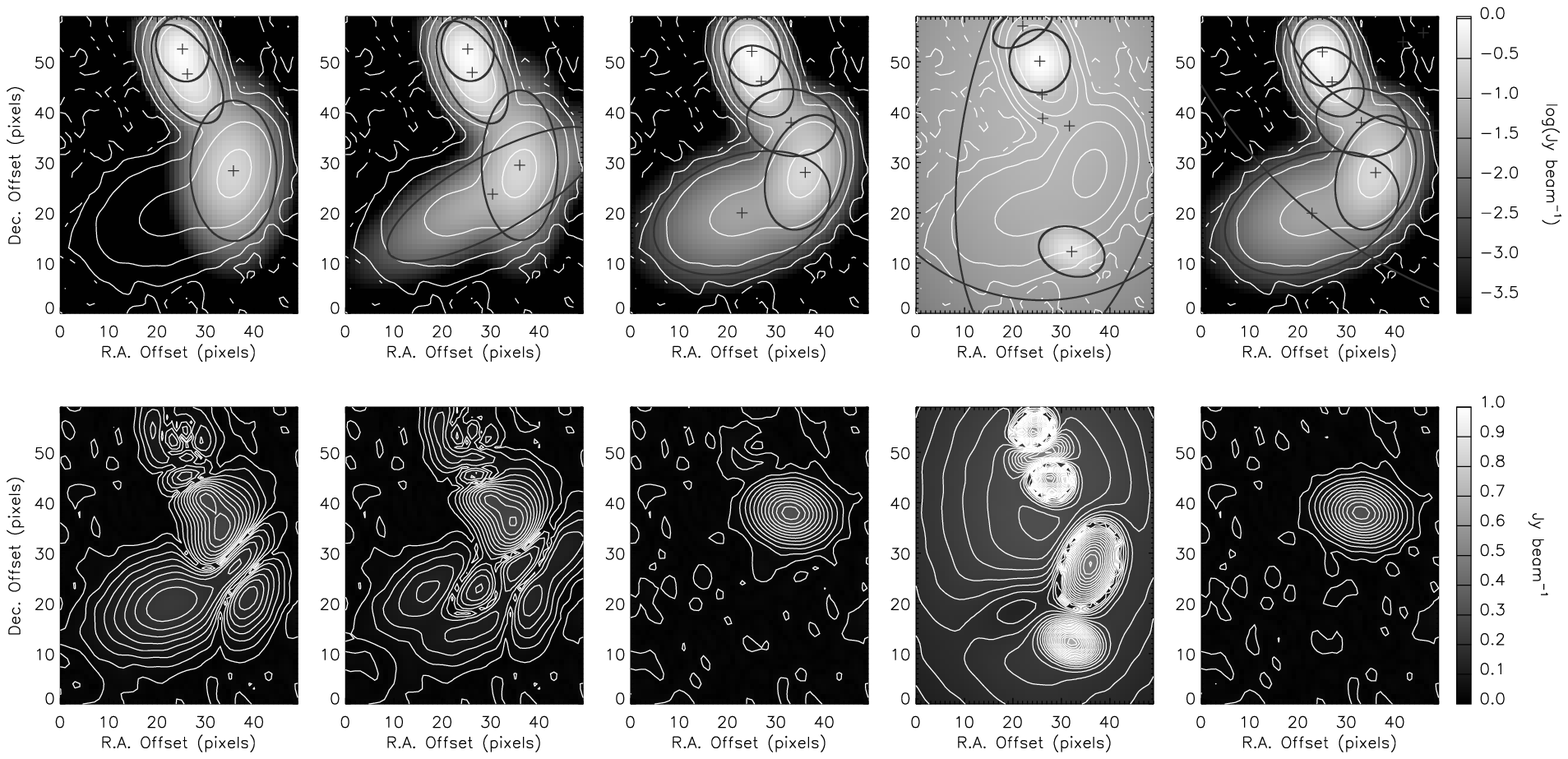}}
          \caption{Results of the model fitting for the synthetic image of jet J2 with addition of a realistic noise pattern obtained after varying the number of sources from two to seven (left-to-right direction). $\it Top:$ Contour lines are of the original image shown in Figure \ref{noise_plus_map_J2}, the gray-scale image is constructed from the fitted source parameters and the dark ellipses are the contours of the individual fitted sources (respective centers marked with crosses) at the FWHM. $\it Bottom:$ Respective residual maps in linear scale.}
      \label{img_noise_synth_J2}
   \end{figure*}

The CE model-fitting results are shown in Figures \ref{img_noise_synth_J1} and \ref{img_noise_synth_J2}. We note that the choice $N_\mathrm{s}=3-6$ for jet J1 and $N_\mathrm{s}=4, 5$, and 7 for jet J2 provides more homogeneous and less peaked residual maps.

\begin{figure}
     \epsscale{0.25}

	  {\plotone{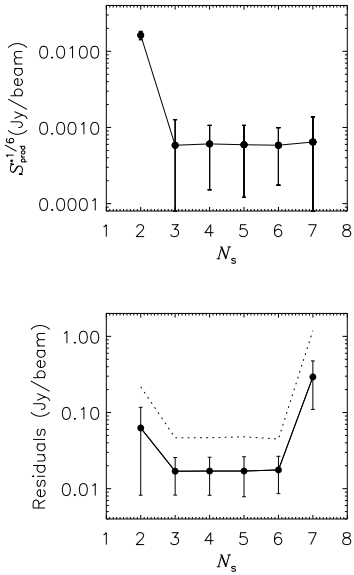}}
          \caption{Quantitative analyses of the model fittings in the case of the synthetic noise image of jet J1. {\it Left:} One-sixth of the performance function as a function of the number of sources used in the CE optimization. Taking into account the error bars, there is a plateau-like structure for $N_\mathrm{s}\ge 3$. {\it Right:} Behavior of the mean value of the residual image shown in Figure \ref{img_noise_synth_J1} in terms of the sources used in the CE optimization. The error bars correspond to the standard deviation of the residual images. The dotted line refers to the maximum value of the residuals.}
      \label{fitness_img_noise_J1}
   \end{figure}

Considering the uncertainties, there is a plateau after $N_\mathrm{s}=3$ in the $S_\mathrm{prod}^{1/6}$ graph in the top panel of Figure \ref{fitness_img_noise_J1}. This means that only a lower limit for $N_\mathrm{s}$ can be derived from such plots. The bottom panel of Figure \ref{fitness_img_noise_J1} points out that residuals are minimized for $3\le N_\mathrm{s}\le 6$. To break the degeneracy in $N_\mathrm{s}$, we analyze the values listed in Table \ref{tab_J1_modfit_noise}. As in the case of the noiseless images, the optimized parameters of the three sources found for $N_\mathrm{s}=3$ also appear when we assumed $4\le N_\mathrm{s}\le 6$. The extra components are dimmer and sometimes superpose completely with the brightest component (core). Thus, $N_\mathrm{s}=3$ is fully favored in relation to the other possibilities.

\begin{figure}
     \epsscale{0.25}

	  {\plotone{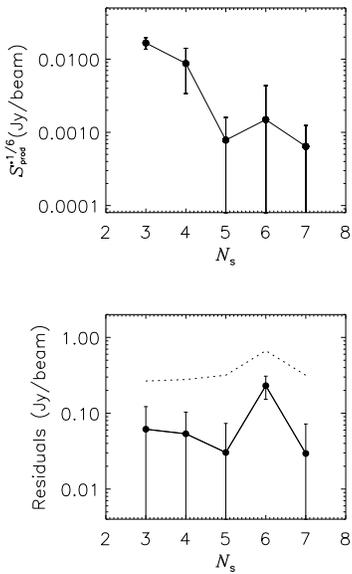}}
          \caption{Quantitative analyses of the model fittings in the case of the synthetic noise image of jet J2. {\it Left:} One-sixth of the performance function as a function of the number of sources used in the CE optimization. {\it Right:} Behavior of the mean value of the residual image shown in Figure \ref{img_noise_synth_J2} in terms of the sources used in the CE optimization. The error bars correspond to the standard deviation of the residual images. The dotted line refers to the maximum value of the residuals.}
      \label{fitness_img_noise_J2}
   \end{figure}

In the case of jet J2, there is also a plateau-like feature for $N_\mathrm{s}\gtrsim 5$ in $S_\mathrm{prod}^{1/6}$ graph in the top panel of Figure \ref{fitness_img_noise_J2}. Solutions with $N_\mathrm{s}=5$ and 7 are slightly favored over the others in the case of the mean and maximum residuals shown in the bottom panel of Figure \ref{fitness_img_noise_J2}. The combination of both results suggests the existence of five or seven components in jet J2. Comparison between the values listed in Table \ref{tab_J2_modfit_noise} reveals that components 1-5 for $N_\mathrm{s}=5$ were also found in the CE optimization using $N_\mathrm{s}=7$ (components 1, 2, 3, 6, and 7, respectively). As in the case of the noisy jet J1, the extra two components are too dim, compatible with null intensity considering the uncertainties. Thus, a five-components solution is favored, as expected for this benchmark test.

\subsection{Checking the Auto-consistency of the Error Estimation of the CE Model-fitting Parameters}

The equations to calculate the model-fitting parameters, as well as their respective uncertainties, were introduced in Section 2.3. To show that Equations (10) and (11) provide a reasonable estimation of the model parameters and their errors, we used the three optimization runs performed for the noisy jet J1 for $N_\mathrm{s}=3$. The reason for not choosing a noiseless image is to keep this analysis as close as possible to real situations, which we know to have a structured noise pattern embedded into the source signal.

\begin{figure*}
     \epsscale{0.8}

	  {\plotone{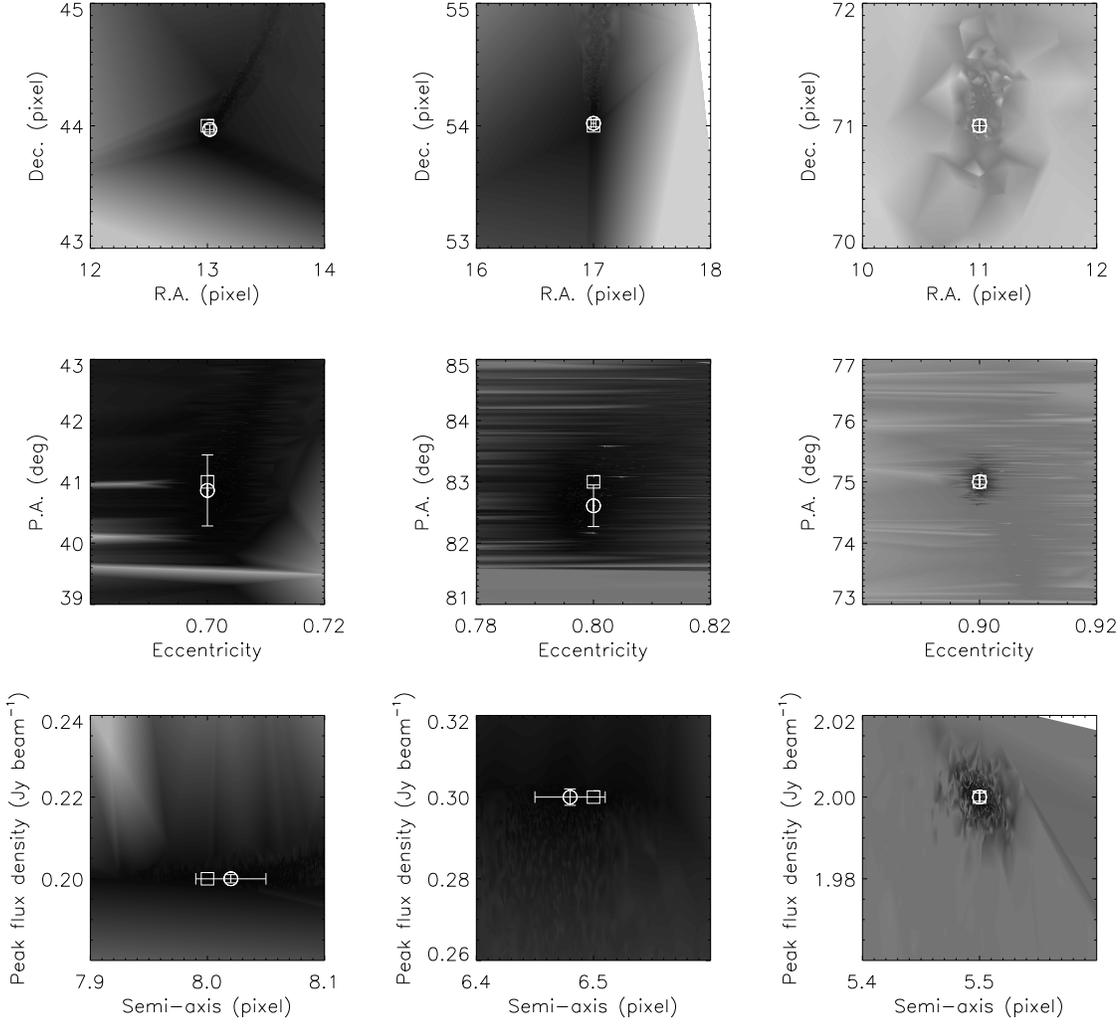}}
          \caption{Mapping of the logarithm distribution of $S_\mathrm{prod}$ in terms of the CE elliptical Gaussian parameters for components 3, 2, and 1 (rightmost, middle, and leftmost panels, respectively), represented by the gray scale (the lower the value of $S_\mathrm{prod}$, the darker the color). Square symbols mark the parameter values listed in Table \ref{tab_J1}, while open circles represent the CE model fitting parameters of the three-source entries in Table \ref{tab_J1_modfit_noise}. Error bars correspond to the $3\sigma$ uncertainties. Note that the CE values are in agreement with the real ones and both are located at the deepest minimum of the objective function in each image.}
      \label{mapping_Nelite}
   \end{figure*}

From the 18,000 best-value data generated in this test (three optimization runs, each one composed of 6000 iterations), we determined the distribution of the logarithm of $S_\mathrm{prod}$ in the ($x,y$)-, ($\epsilon,\theta$)-, and ($a,I_0$)-planes for each of the three Gaussian sources, as is displayed in Figure \ref{mapping_Nelite}\footnote{This representation of $S_\mathrm{prod}$ was chosen due to its complex functional form, which depends on 18 parameters for this particular benchmark test.}. Although the $S_\mathrm{prod}$ distribution is very irregular, showing several local minima in all planes, the CE algorithm was able to find the global minimum of $S_\mathrm{prod}$, returning the correct parameter values of the ellipses if the error bars were considered. For example, the differences between the $x-$ and $y-$values listed in Table \ref{tab_J1} and those calculated from our technique are much smaller than one pixel in both directions. The larger discrepancies were found in the $(\epsilon,\theta)-$planes but, even in these cases, such differences remain within the $3\sigma$ errors.

Our results show also that it is possible to calculate statistically the values of the model-fitting parameters and their associated uncertainties by applying Equations (10) and (11) to the tentative solutions generated in the optimization process. However, as the real nature of the noise present in interferometric images is not sufficiently known, such error estimates should be considered as a lower limit for the true uncertainties of the derived Gaussian parameters. Error estimations based on the properties of the restoring beam or differences in the fits to two or more observations acquired very close in time (e.g., \citealt{pin07,lis09b}) may be used as complements or alternatives to Equations (10) and (11) in general situations.

\subsection{The Influence of S/R on the CE Model-fitting Optimization Process}

\begin{figure}
     \epsscale{0.3}
	  {\plotone{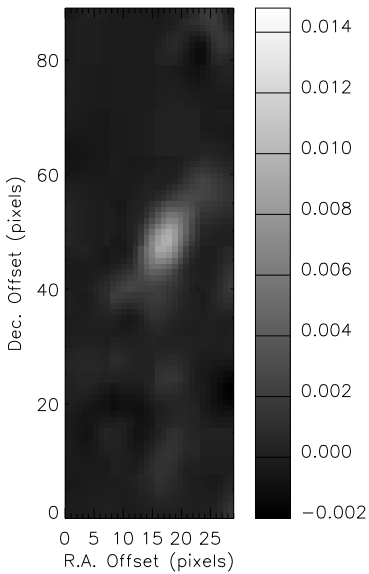}}
          \caption{Image of a synthetic elliptical Gaussian component superposed on the background noise shown in the left panel of Figure \ref{noise_plus_map_J1} with S/N equal to 30. The gray-scale map as well as the contours is in linear scale in units of Jy beam$^{-1}$.}
      \label{im_1gauss_SNR30}
   \end{figure}

The images in the noiseless validation tests had a formal infinite S/N, while for the J1 and J2 noise images S/R was $\sim 100$. To check the performance of our CE model-fitting algorithm in terms of S/R, we created synthetic images composed of the same noise background shown in the top panel of Figure \ref{noise_plus_map_J1} and an elliptical Gaussian source centered at $x=17.2$ pixels and $y=48.1$ pixels. We placed this source at this particular position in order to superpose it on a region in which there is a strong gradient in the noise intensity in the southeast to northwest direction, which introduces extra difficulty to the model-fitting procedure. The additional parameters were chosen to be $a=4.8$ pixels, $\epsilon=0.85$, and $\theta=67\fdg 3$. The peak intensity of this source was varied from 1 to 50 mJy beam$^{-1}$, which corresponds to S/R ranging from 2 to 100. An example of one of these control images is displayed in Figure \ref{im_1gauss_SNR30}.

\begin{figure*}
     \epsscale{0.8}

	  {\plotone{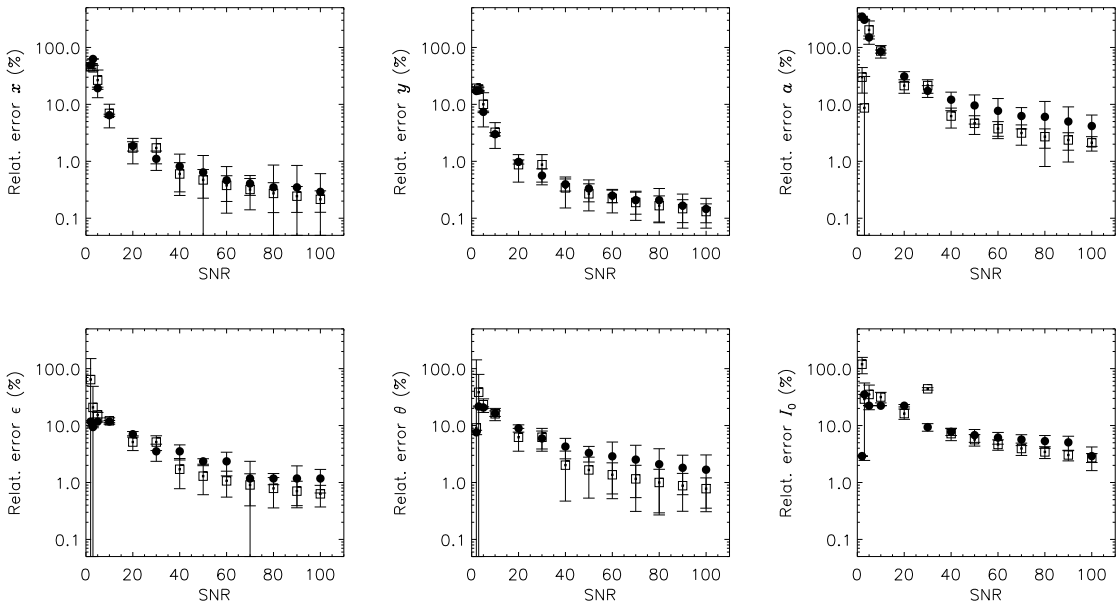}}
          \caption{Relative errors in the parameters of an elliptical Gaussian source as a function of the S/N. The black circles correspond to the results obtained from our CE model fitting technique, while the open squares refer to the model fitting obtained from the AIPS task IMFIT.}
      \label{onegaussSNR}
   \end{figure*}

The efficiency of our algorithm in recovering the elliptical Gaussian source parameters as a function of the S/R can be seen in Figure \ref{onegaussSNR}. The values of the Gaussian-peak coordinates $x$ and $y$ were recovered with a relative error below the 10$\%$ level for S/R$\gtrsim 10$. The same error level was achieved for the parameters $\epsilon$ and $\theta$ for S/R$\gtrsim 20$, $I_0$ for S/R$\gtrsim 30$, and $a$ for S/R$\gtrsim 50$. For SNR = 100 all parameters were found with a relative error below the 5$\%$ level. 

In terms of the absolute values and their respective uncertainties, our CE model fitting technique was able to recover the parameter $I_0$ within the $3\sigma$ level for S/R$\gtrsim 20$, while for the other parameters the same applied for S/R$\gtrsim 10$.

\subsection{Comparing the CE Benchmark Results with Those Obtained from the AIPS Task IMFIT}

The aim of this subsection is to compare our findings with those obtained from the traditional and widely used two-dimensional Gaussian fitting task IMFIT included in the Astronomical Image Processing System Package (AIPS).

A comparison between CE and IMFIT model-fitting results in the case of the S/R limit tests is shown in Figure \ref{onegaussSNR}. Taking into account the uncertainties, both techniques have very similar behavior in terms of recovering the source parameters for this particular test.

We applied the task IMFIT to the image of the noisy jet J1 displayed in Figure \ref{noise_plus_map_J1}. A comparison between the values of the parameters of the three Gaussian sources of J1 from IMFIT and from our CE technique is shown in Table \ref{J1_CE_IMFIT}. Again, both algorithms found the Gaussian parameters with similar high efficiency.

Unfortunately, IMFIT cannot fit simultaneously more than four Gaussian sources. To make possible a comparison between those algorithms, we eliminate source number 3 of the noisy jet J2, transforming it into a four-component image labeled as J3\footnote{The IMFIT failed to fit four components in the original image of the noisy jet J2, displaying an error message related to some convergence issue. Using three components, IMFIT returns parameters very similar to those listed in the three-components entry of the Table \ref{tab_J2_modfit_noise}}. This makes J3 map as a set of two overlapping components with S/R $\sim 100$. We show in Table \ref{J3_CE_IMFIT} the model-fitting values recovered from the CE and IMFIT fittings as a function of the correct ones. Unlike the earlier comparisons presented in this subsection, the CE technique had a superior performance in determining the structural parameters of the Gaussian sources, especially in the case of the dimmest component for which IMFIT failed to converge to the correct parameters. In terms of relative errors, our CE algorithm was able to recover 7 out of 24 parameters within a zero-percent level, and only one parameter with an error higher than 5 $\%$ level (more precisely, the 6.8$\%$ level for the parameter $\theta$ of the dimmest source). On the other hand, the relative errors obtained from IMFIT are systematically higher in 19 out of 24 parameters. In the case of the dimmest source, the relative errors ranged from 16$\%$ to 400$\%$, indicating that the IMFIT did not converge for this component.

In summary, comparison between the CE optimization algorithm and IMFIT suggests that both tools have similar performance in dealing with images that are not too complex (few and relatively well-separated sources). However, our technique has a better performance when the interferometric map is more complex, as in the case of test J3. Note also that our method does not have a limit for the number of components to be fitted simultaneously in the image, as in the case of IMFIT.

\section{Application of the CE technique to a real interferometric image}
 
We have presented in previous sections some validation tests necessary to demonstrate the capability of our model-fitting technique of dealing with interferometric images. The objective of this section is to check the behavior of our algorithm when applied to real interferometric images. To do this, we selected a naturally weighted $I-$image taken from the MOJAVE/2cm Survey Data Archive \citep{lis09a} that corresponds to the 15 GHz radio map of the BL Lac OJ\,287 obtained in 1996 May 27.

\subsection{CE Model-fitting and the Number of Jet Components}

The original fits image of the OJ\,287 is formed by an array of 512 $\times$ 512 pixels but only a relatively small fraction has a jet signal significantly higher than the noise. Because of this, we decided to crop the original fits image to maintain only the fraction with a useful signal, which meant a 66 $\times$ 51 pixel image. It is important to emphasize that this reduction helps the algorithm to find the optimal solutions more rapidly since the parameter space is substantially narrowed in this case. 

Several works in the literature have assumed a circular Gaussian shape for jet features (e.g., \citealt{lob01,jor05,agu07}). In this work, we have assumed a two-dimensional elliptical Gaussian function as being representative of the brightness distribution of the jet knots, which will be kept for the radio map of the BL Lac OJ\,287. The justification for this is that our CE model fitting will be executed in the image plane, in which the jet components are convolved with the elliptical synthesized CLEAN beam of the interferometric experiment.

We applied our CE algorithm to the cropped image, varying the number of elliptical Gaussian sources from two to seven. For each adopted number of sources we run the algorithm three times. The model fitting processes examined the whole image in the $x-$ and $y-$ coordinates, searching for Gaussian sources with structural parameters in the following ranges: $4.0\leq a \mathrm{(pixel)}\leq 30$, $0\leq \epsilon\leq 0.9$, $-90\degr\leq \theta \leq 90\degr$ (positive values for north to east direction), and $9.51\times 10^{-4}\leq I_0$ (Jy beam$^{-1}$)$\leq 1.2$. The lower value of $I_0$ corresponded to twice the nominal root mean square of the OJ\,287 image.

\begin{figure*}
     \epsscale{0.53}

	  {\plotone{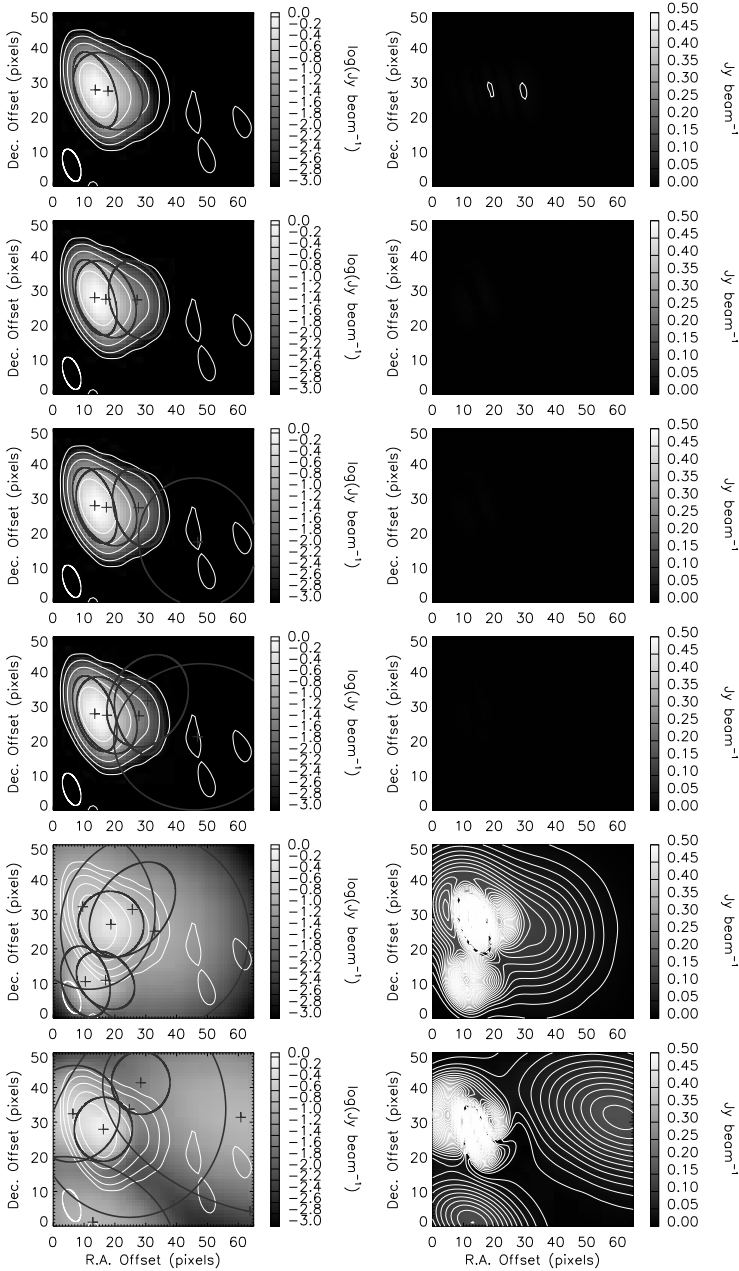}}
          \caption{Model fittings in the case of 15 GHz VLBA image of the BL Lac OJ\,287 using elliptical Gaussian sources. $\it Left:$ White contour lines represent the VLBA map of OJ\,287 in 1996 May 27, the gray-scale image is constructed from the fitted source parameters and the dark thick ellipses are the contours of the individual fitted sources at the FWHM (center marked with crosses). The number of fitted Gaussian sources increases from the top to the bottom panel (from two to seven sources, respectively). The thin white ellipses at the lower leftmost corner in each panel are the FWHM synthesized beam of the observation. $\it Right:$ Residual maps in linear scale. The lower left corner in all panels, i.e. the origin of the right ascension and declination offset coordinates, represents pixels 240 and 230 of the original fits image of OJ\,287 in right ascension and declination directions, respectively.}
      \label{OJ287_960527_ellip}
   \end{figure*}

\begin{figure}
     \epsscale{0.25}

	  {\plotone{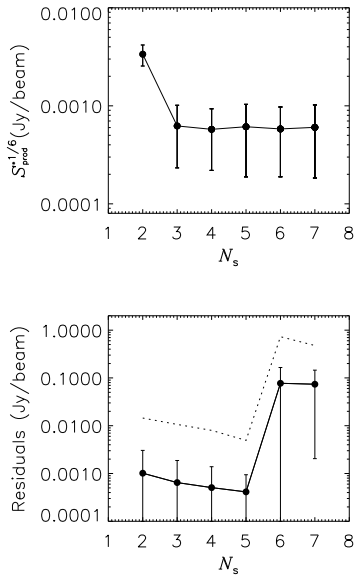}}
          \caption{Quantitative analyses of the model fittings in the case of 15 GHz VLBA image of the BL Lac OJ 287 in 1996 May 27 using elliptical Gaussian sources. {\it Top:} One-sixth of the objective function as a function of the number of the sources used in the CE optimization. Note the plateau-like structure (considering the error bars) for $N_\mathrm{s}\ge 3$. {\it Bottom:} Behavior of the mean value of the residual image shown in the lower panels of the Figure \ref{OJ287_960527_ellip} in terms of the sources used in the CE optimization. The error bars correspond to the standard deviation of the residual image. The dotted line refers to the maximum value of the residuals.}
      \label{fitness_OJ287_960527_ellip}
   \end{figure}

The optimal values for the model parameters and their respective uncertainties were estimated following Section 2.3 and are displayed in Table \ref{tab_OJ287_modfit_ellipt}. The CE optimization results can be seen in Figure \ref{OJ287_960527_ellip}, in which  the Gaussian components are shown superposed on the observed image, as well as the residual maps. A visual inspection of the residual maps reveals less-structured residuals when three to five sources are employed in the optimization. The top panel of Figure \ref{fitness_OJ287_960527_ellip} shows that the performance function introduced in Section 2.2 is better minimized by adopting $N_\mathrm{s}\gtrsim 3$. The behavior of the mean and maximum values of the residual image shown in the bottom panel of Figure \ref{fitness_OJ287_960527_ellip} seems to be minimized for  $3\lesssim N_\mathrm{s}\lesssim 5$. Therefore, the use of Figures \ref{OJ287_960527_ellip} and \ref{fitness_OJ287_960527_ellip} only provides an optimal range for the number of components present in the image of OJ\,287. However, looking carefully at the entries of Table \ref{tab_OJ287_modfit_ellipt}, we realize that components 1, 2, and 3 for $N_\mathrm{s}=3$ were also found in the CE optimizations using $N_\mathrm{s}=4$ and 5. The extra components in both cases are too dim ($\sim 1$ mJy beam$^{-1}$), compatible with the zero-intensity level if uncertainties are considered. Besides, similar plots of those shown in Figure \ref{Param_converg_J2} reveal that those extra components introduced large oscillations in the values of the Gaussian parameters. We believe that such findings point firmly to the existence of three components in the jet of OJ\,287 in 1996 May 27.

\subsection{Comparing the CE Results with Those in the Literature}

As discussed in the previous section, our results seem to indicate the presence of three components in the jet of OJ\,287 (core plus two components). Interestingly, \citet{lis09b} also assumed the existence of three sources in the 15 GHz image of OJ\,287 in 1996 May 27. Their modeling results were obtained by adopting circular-shaped sources fitted in the interferometric $(u,v)$ plane instead of in the image plane, as has been done in this work. A comparison between our model-fitting parameter values for $N_\mathrm{s}=3$ and those found by \citet{lis09b} can be seen in Table \ref{table_comparative_OJ287_960527_v2}. 

To construct Table \ref{table_comparative_OJ287_960527_v2}, we transformed our original optimized data listed in Table \ref{tab_OJ287_modfit_ellipt} to the format given in Table 1 of \citet{lis09b}. The first parameter is the core-component distance $r$ that can be calculated from

\begin{equation}
r=\sqrt{(x-x_\mathrm{core})^2+(y-y_\mathrm{core})^2},
\end{equation}
where $x_\mathrm{core}$ and $y_\mathrm{core}$ are, respectively, the right ascension and declination coordinates of component 1 in Table \ref{tab_OJ287_modfit_ellipt}.

The position angles of the components measured in relation to the core component, $\eta$, taken from \citet{lis09b} were subtracted by $360\degr$ before being displayed in Table \ref{table_comparative_OJ287_960527_v2}. 

As mentioned previously, we fitted elliptical Gaussian sources in the image plane. \citet{lis09b} adjusted circular Gaussian sources to the $(u,v)$ data, which allowed them to obtain for some components sizes smaller than that of the restoring beam. Remembering that the convolution of two Gaussian functions results in a Gaussian function with a squared full width at half-maximum (FWHM) proportional to their FWHM, we determined the effective size of the optimized components $a_\mathrm{FWHM}$ through:

\begin{equation}
a_\mathrm{FWHM}=\sqrt{a^2\left(1-\epsilon^2\right)^{1/2}-b_\mathrm{FWHM}^\mathrm{maj}b_\mathrm{FWHM}^\mathrm{min}},
\end{equation}
where $b_\mathrm{FWHM}^\mathrm{maj}$ and $b_\mathrm{FWHM}^\mathrm{min}$ are respectively the FWHM of the major and minor synthesized beam axes for the image epoch (1.16 $\times$ 0.60 mas; \citealt{lis09b}). 

The flux density $F$ of the jet components can be estimated from:

\begin{equation}
F=8\ln{2}\left(\frac{a^2\sqrt{1-\epsilon^2}}{b_\mathrm{FWHM}^\mathrm{maj}b_\mathrm{FWHM}^\mathrm{min}}\right)I_0.
\end{equation}

In the last equation, $a$, $b_\mathrm{FWHM}^\mathrm{maj}$ and $b_\mathrm{FWHM}^\mathrm{min}$ must be given in pixels, and $I_0$ in Jy beam$^{-1}$ in order to have $F$ in units of Jy.

We note that there is a relative good agreement between our results and those found by \citet{lis09b} if we take into account the uncertainties. It suggests that our CE technique may provide results as robust as those model fittings performed in the $(u,v)$ plane. Further studies using a larger set of real images are required to confirm undoubtedly this claim.

The optimized Gaussian sources shown in Figure \ref{OJ287_960527_ellip} represent the convolution between the source itself and the synthesized beam of the observation. If the source is punctual, such a convolution must return the shape of the synthesized beam. For the image of OJ\,287 considered in this work, the beam is elliptical, with its major and the minor semi-axes being respectively 4.91 and 2.56 pixels (or 0.58 and 0.30 mas) in size, and with a position angle $P.A.=22\fdg 28$ on the plane of the sky \citep{lis09b}. This last quantity is related to our definition of $\theta$ through $P.Aa.=90\degr+\theta$. A comparison of these values with those listed in Table \ref{tab_OJ287_modfit_ellipt} for $N_\mathrm{s}=3$ shows that component 1 is unresolved by the observation, presenting major and minor semi-axes equals respectively to 4.92 and 2.59 pixels, as well as $P.A.=22\fdg 70$, in full agreement with what is expected from a punctual source, which is indeed the compact core of OJ\,287. This result further reinforces the great potential of our technique in modeling interferometric radio images.

\section{Conclusion}

We have developed a new method to obtain model fittings to interferometric radio images of astrophysical jets using a global optimization algorithm known as cross-entropy. To validate our model-fitting optimized procedure, we built benchmark tests that employed synthetic images created to simulate as realistically as possible typical real interferometric radio maps. We assumed that the individual sources of a jet can be represented by two-dimensional Gaussian functions defined by six parameters: peak intensity coordinates, peak intensity, size and angle of the semi-major axis, and eccentricity. 

The first validation tests optimize synthetic images with infinite S/R composed of three and five sources (jets J1 and J2, respectively). The parameters for each source were selected in an attempt to reproduce typical image characteristics encountered in VLBI AGN jet images. The results of our fitting technique for these synthetic images were excellent: all parameter values were recovered by our technique taking into account the formal uncertainties. The relative errors of these two benchmark tests are null in most cases, and do not exceed 0.25\% in the worst cases.

The second set of benchmark tests was built from jets J1 and J2, embedding them in a structured noise backgrounds extracted from typical MOJAVE images. Our CE model-fitting technique was able to recover the parameters of the three sources of the noisy jet J1 in most cases with null relative error. In the case of the complex noisy jet J2, 18 out of 30 parameters had values below of 0.3\% in terms of relative errors (seven among them having a zero-percent value).

Plots of the values of the performance function and the mean and maximum amplitude of the residual images as a function of the number of tentative components only provided an optimal range for $N_\mathrm{s}$ in the majority of the benchmark tests presented in this work. However, we showed that a careful analysis of the optimized parameters among all used values of $N_\mathrm{s}$ can remove such degeneracy, and determine the true value of the number of components. If the assumed $N_\mathrm{s}$ is smaller than the real one, our CE optimization tries to fit the brightest components since their contribution must be preponderant in the minimization of the residuals. On the other hand, if the adopted $N_\mathrm{s}$ is larger than the correct one, the extra components tends to be too dim (compatible with null intensity if uncertainties are considered) or they are practically coincident with the most intense sources in the image. Thus, the true jet components will always be present in the optimized images when we assumed a number of sources greater than the real one.

The number of components also has an influence on the convergence of the parameters: the inclusion of more components than necessary introduces an oscillatory pattern, as well as some strong discontinuities in the plots of the values of the Gaussian parameters as a function of the iteration number. This can also be used to constrain the value of $N_\mathrm{s}$.

The cross-entropy technique is able to recover the parameters of the sources with a similar accuracy to that obtained from the traditional AIPS task IMFIT when the image is relatively simple (e.g., few components). For more complex interferometric maps, our method exhibits a superior performance in recovering the parameters of the jet components. 

To verify the performance of our algorithm in the case of real observational data, we applied it to the image of the BL Lac OJ\,287 obtained in 1996 May 27 by the MOJAVE consortium, as mentioned in Section 4. The behavior of the objective function, as well as the mean and the maximum values of the residual maps as a function of the number of the sources, indicates the presence of three to five components. This degeneracy is broken when the data of Table \ref{tab_OJ287_modfit_ellipt} are carefully analyzed. They indicate the presence of the three sources found in the optimization with $N_\mathrm{s}=3$ in the results from $N_\mathrm{s}=4$ and 5, with the extra components too dim to be taken seriously. It is important to emphasize that the structural parameters of the core found by our technique are in full agreement with those expected from the convolution of a point-like source and an elliptical restoring beam. Interestingly, \citet{lis09b} modeled the same image of OJ\,287 in the $(u,v)$ plane as the composition of three circular Gaussian sources. The structural source parameters obtained by these authors are in good agreement with those found in this work from our CE optimization technique. This suggests that our image-based model-fitting method might be as efficient as $(u,v)$-based modeling despite the usual deconvolution and bad $(u,v)$ coverage artifacts that might be present in CLEAN images. The next step in our tests will be to adapt the CE method to fittings in the $(u,v)$ plane.

It is important to point out that we have assumed throughout this work that the components can be modeled by two-dimensional Gaussian functions, which might not be suitable in some situations (e.g., in pronounced bow-shock regions usually seen in terminal jets). Indeed, there is no prior guarantee that elliptical Gaussian functions can always represent the correct shape of jet knots in general.

We believe that our results indicate that this new optimization technique can provide a major contribution in obtaining model fittings to VLBI images of astrophysical jets. In particular, our CE model-fitting technique must be used in studies of complex emission regions presenting more than three sources, even though it is substantially slower than current model-fitting tasks (at 10,000 times slower for a single processor, depending on the number of sources to be optimized). As in the case of any model fitting performed in the image plane, caution is always required in analyzing images constructed from a poorly sampled $(u,v)$ plane.

\acknowledgments

This work was supported by the Brazilian Agencies FAPESP (Proc. 2006/57824-1) and CNPq. This research has made use of data from the MOJAVE database that is maintained by the MOJAVE team (Lister et al., 2009a). The authors
acknowledge very helpful remarks from an anonymous referee.

\clearpage


\begin{table}
\footnotesize
\begin{center}
  \caption{Parameters of the Three Gaussian Sources in the Case of the Jet J1, Shown in Figure \ref{im_3gauss_no_noise}. \label{tab_J1}}
  \begin{tabular}{@{}ccccccc@{}}
\tableline\tableline
   & $x$ & $y$ & $a$ & $\epsilon$ & $\theta$ & $I_0$\\
& (pixel) & (pixel) & (pixel) & & (deg) & (Jy/beam) \\
 \hline
1 & 11.0 & 71.0  &    5.5  &   0.90  &    75.0   &  2.0\\
2 & 17.0 & 54.0  &    6.5  &   0.80  &    83.0   &  0.3\\
3 & 13.0 & 44.0  &    8.0  &   0.70  &    41.0   &  0.2\\
\tableline
\end{tabular}
\tablecomments{Component 1 represents the core, while components 2 and 3 are the jet knots.}
\end{center}
\end{table}



\begin{table}
\footnotesize
\begin{center}
  \caption{Parameters of the Five Gaussian Sources in the Case of the Jet J2, Shown in Figure \ref{im_5gauss_no_noise}. \label{tab_J2}}
  \begin{tabular}{@{}ccccccc@{}}

  \tableline\tableline
   & $x$ & $y$ & $a$ & $\epsilon$ & $\theta$ & $I_0$\\
& (pixel) & (pixel) & (pixel) & & (deg) & (Jy beam$^{-1}$) \\
\tableline
1 & 25.0 & 52.0  &    3.0  &   0.60  &   -70.0   &  1.00\\
2 & 27.0 & 46.0  &    3.2  &   0.65  &   -50.0   &  0.50\\
3 & 33.0 & 38.0  &    4.0  &   0.70  &   -10.0   &  0.10\\
4 & 36.0 & 28.0  &    5.0  &   0.75  &    70.0   &  0.30\\
5 & 23.0 & 20.0  &    8.0  &   0.80  &    20.0   &  0.05\\
\tableline
\end{tabular}
\tablecomments{Component 1 represents the core, while components 2-5 are the jet knots.}
\end{center}
\end{table}



\begin{table*}
\footnotesize
\begin{center}
  \caption{Model Fitting Parameters of the Gaussian Sources in the Case of the  Noiseless Synthetic Image of J1. \label{tab_J1_modfit_no_noise}}
  \begin{tabular}{@{}ccccccc@{}}
\tableline\tableline
   & $x$ & $y$ & $a$ & $\epsilon$ & $\theta$ & $I_0$ \\
& (pixel) & (pixel) & (pixel) & & (deg) & (Jy/beam) \\
\tableline
    1      &        11.00  $\pm$         0.01   &       71.02  $\pm$         0.02   &        5.47  $\pm$         0.02   &        0.90  $\pm$         0.00   &       74.96  $\pm$         0.16   &       2.003  $\pm$        0.006  \\
    2      &        15.56  $\pm$         0.04   &       49.55  $\pm$         0.06   &        9.21  $\pm$         0.07   &        0.85  $\pm$         0.00   &       69.39  $\pm$         0.48   &       0.368  $\pm$        0.002  \\\\
    1      &        11.00  $\pm$         0.00   &       71.00  $\pm$         0.00   &        5.50  $\pm$         0.00   &        0.90  $\pm$         0.00   &       75.00  $\pm$         0.01   &       2.000  $\pm$        0.001  \\
    2      &        17.00  $\pm$         0.01   &       54.00  $\pm$         0.02   &        6.50  $\pm$         0.01   &        0.80  $\pm$         0.00   &       83.00  $\pm$         0.17   &       0.300  $\pm$        0.001  \\
    3      &        13.00  $\pm$         0.02   &       44.00  $\pm$         0.02   &        8.00  $\pm$         0.02   &        0.70  $\pm$         0.00   &       40.99  $\pm$         0.39   &       0.200  $\pm$        0.001  \\\\
    1      &        11.00  $\pm$         0.00   &       71.00  $\pm$         0.00   &        5.50  $\pm$         0.00   &        0.90  $\pm$         0.00   &       75.00  $\pm$         0.02   &       2.000  $\pm$        0.001  \\
    2      &        17.00  $\pm$         0.01   &       54.00  $\pm$         0.02   &        6.50  $\pm$         0.02   &        0.80  $\pm$         0.00   &       83.00  $\pm$         0.20   &       0.299  $\pm$        0.001  \\
    3      &        17.54  $\pm$         4.74   &       52.96  $\pm$         9.80   &        5.13  $\pm$         2.73   &        0.54  $\pm$         0.43   &       40.02  $\pm$       102.94   &       0.001  $\pm$        0.001  \\
    4      &        13.00  $\pm$         0.03   &       44.00  $\pm$         0.04   &        8.00  $\pm$         0.03   &        0.70  $\pm$         0.00   &       41.02  $\pm$         0.43   &       0.200  $\pm$        0.001  \\\\
    1      &        23.88  $\pm$         0.32   &       87.79  $\pm$         0.23   &       22.97  $\pm$         0.55   &        0.94  $\pm$         0.01   &       89.00  $\pm$         1.54   &       0.000  $\pm$        0.001  \\
    2      &        26.69  $\pm$         1.57   &       86.26  $\pm$         2.24   &       37.08  $\pm$         2.65   &        0.92  $\pm$         0.04   &       80.14  $\pm$        10.76   &       0.000  $\pm$        0.001  \\
    3      &        11.00  $\pm$         0.00   &       71.00  $\pm$         0.01   &        5.50  $\pm$         0.01   &        0.90  $\pm$         0.00   &       75.00  $\pm$         0.07   &       2.000  $\pm$        0.003  \\
    4      &        17.00  $\pm$         0.04   &       54.00  $\pm$         0.09   &        6.49  $\pm$         0.07   &        0.80  $\pm$         0.01   &       83.14  $\pm$         0.82   &       0.300  $\pm$        0.004  \\
    5      &        13.00  $\pm$         0.12   &       44.01  $\pm$         0.13   &        7.98  $\pm$         0.09   &        0.70  $\pm$         0.01   &       41.10  $\pm$         1.61   &       0.200  $\pm$        0.003  \\\\
    1      &        12.93  $\pm$         4.78   &       78.87  $\pm$         9.04   &        3.41  $\pm$         1.43   &        0.63  $\pm$         0.44   &       19.44  $\pm$        91.56   &       0.001  $\pm$        0.001  \\
    2      &        11.00  $\pm$         0.00   &       71.00  $\pm$         0.00   &        5.50  $\pm$         0.00   &        0.90  $\pm$         0.00   &       75.00  $\pm$         0.03   &       2.000  $\pm$        0.001  \\
    3      &        10.16  $\pm$         3.87   &       68.13  $\pm$         6.75   &        3.44  $\pm$         1.48   &        0.52  $\pm$         0.49   &       30.23  $\pm$        80.18   &       0.001  $\pm$        0.001  \\
    4      &         9.15  $\pm$         2.44   &       63.30  $\pm$         3.90   &        3.21  $\pm$         1.10   &        0.64  $\pm$         0.42   &       45.78  $\pm$        62.45   &       0.001  $\pm$        0.001  \\
    5      &        17.00  $\pm$         0.01   &       54.00  $\pm$         0.03   &        6.50  $\pm$         0.02   &        0.80  $\pm$         0.00   &       83.00  $\pm$         0.27   &       0.300  $\pm$        0.001  \\
    6      &        13.00  $\pm$         0.04   &       44.00  $\pm$         0.04   &        8.00  $\pm$         0.03   &        0.70  $\pm$         0.00   &       40.99  $\pm$         0.55   &       0.200  $\pm$        0.001  \\\\
    1      &         9.45  $\pm$         0.01   &       69.86  $\pm$         0.00   &        2.81  $\pm$         0.01   &        0.00  $\pm$         0.00   &       -6.75  $\pm$         0.29   &       0.739  $\pm$        0.005  \\
    2      &         3.11  $\pm$         0.05   &       65.92  $\pm$         0.04   &        4.14  $\pm$         0.01   &        0.95  $\pm$         0.00   &       51.00  $\pm$         0.11   &       0.000  $\pm$        0.001  \\
    3      &        12.33  $\pm$         0.02   &       64.20  $\pm$         0.05   &       10.81  $\pm$         0.02   &        0.76  $\pm$         0.00   &       57.46  $\pm$         0.22   &       0.717  $\pm$        0.003  \\
    4      &         1.00  $\pm$         0.06   &       56.16  $\pm$         0.03   &        5.40  $\pm$         0.00   &        0.33  $\pm$         0.00   &       -1.58  $\pm$         0.11   &       0.343  $\pm$        0.001  \\
    5      &        11.62  $\pm$         0.03   &       45.71  $\pm$         0.02   &       29.57  $\pm$         0.09   &        0.21  $\pm$         0.00   &       90.00  $\pm$         0.33   &       0.010  $\pm$        0.000  \\
    6      &         1.00  $\pm$         0.03   &       38.16  $\pm$         0.10   &       14.20  $\pm$         0.04   &        0.79  $\pm$         0.00   &       68.25  $\pm$         0.04   &       0.000  $\pm$        0.000  \\
    7      &        16.70  $\pm$         0.01   &       11.98  $\pm$         0.11   &        2.50  $\pm$         0.02   &        0.46  $\pm$         0.00   &      -57.52  $\pm$         0.35   &       0.104  $\pm$        0.000  \\\\

\tableline

\tableline
\end{tabular}
\tablecomments{The number of the sources was varied from two to seven. The uncertainties in each parameter correspond to the $3\sigma$ level.}
\end{center}
\end{table*}



\begin{table*}
\footnotesize
\begin{center}
  \caption{Model Fitting Parameters of the Gaussian Sources in the Case of the  Noiseless Synthetic Image of J2. \label{tab_J2_modfit_no_noise}}
  \begin{tabular}{@{}ccccccc@{}}
\tableline\tableline
   & $x$ & $y$ & $a$ & $\epsilon$ & $\theta$ & $I_0$ \\
& (pixel) & (pixel) & (pixel) & & (deg) & (Jy/beam) \\
\tableline
    1      &        25.24  $\pm$         0.05   &       52.47  $\pm$         0.07   &        2.85  $\pm$         0.11   &        0.62  $\pm$         0.05   &      -72.85  $\pm$         4.00   &       0.731  $\pm$        0.048  \\
    2      &        26.18  $\pm$         0.06   &       47.63  $\pm$         0.28   &        4.84  $\pm$         0.05   &        0.86  $\pm$         0.00   &      -59.23  $\pm$         1.72   &       0.579  $\pm$        0.033  \\
    3      &        35.65  $\pm$         0.06   &       28.33  $\pm$         0.08   &        6.34  $\pm$         0.08   &        0.80  $\pm$         0.01   &       73.32  $\pm$         1.32   &       0.279  $\pm$        0.003  \\\\
    1      &        25.19  $\pm$         0.10   &       52.55  $\pm$         0.24   &        2.75  $\pm$         0.22   &        0.57  $\pm$         0.06   &      -74.18  $\pm$         5.69   &       0.693  $\pm$        0.143  \\
    2      &        26.14  $\pm$         0.39   &       47.85  $\pm$         0.96   &        4.70  $\pm$         0.46   &        0.85  $\pm$         0.05   &      -61.27  $\pm$         2.85   &       0.615  $\pm$        0.080  \\
    3      &        35.93  $\pm$         0.32   &       29.44  $\pm$         0.92   &        6.32  $\pm$         0.29   &        0.85  $\pm$         0.03   &      -89.87  $\pm$         5.30   &       0.234  $\pm$        0.018  \\
    4      &        30.37  $\pm$         1.45   &       23.75  $\pm$         0.85   &       10.36  $\pm$         0.73   &        0.95  $\pm$         0.02   &       28.27  $\pm$         3.41   &       0.077  $\pm$        0.024  \\\\
    1      &        25.00  $\pm$         0.01   &       52.00  $\pm$         0.01   &        3.00  $\pm$         0.01   &        0.60  $\pm$         0.00   &      -70.03  $\pm$         0.46   &       1.000  $\pm$        0.002  \\
    2      &        27.00  $\pm$         0.01   &       46.00  $\pm$         0.02   &        3.20  $\pm$         0.02   &        0.65  $\pm$         0.01   &      -49.98  $\pm$         0.71   &       0.500  $\pm$        0.002  \\
    3      &        32.99  $\pm$         0.09   &       38.00  $\pm$         0.07   &        4.00  $\pm$         0.07   &        0.70  $\pm$         0.02   &      -10.06  $\pm$         2.67   &       0.100  $\pm$        0.002  \\
    4      &        36.00  $\pm$         0.02   &       28.00  $\pm$         0.04   &        5.00  $\pm$         0.03   &        0.75  $\pm$         0.00   &       70.00  $\pm$         0.60   &       0.300  $\pm$        0.002  \\
    5      &        22.97  $\pm$         0.24   &       19.99  $\pm$         0.15   &        7.98  $\pm$         0.21   &        0.80  $\pm$         0.02   &       20.05  $\pm$         2.33   &       0.050  $\pm$        0.001  \\\\
    1      &        18.68  $\pm$         2.38   &       56.72  $\pm$        20.56   &       19.05  $\pm$         6.22   &        0.94  $\pm$         0.08   &      -46.64  $\pm$        37.33   &       0.179  $\pm$        0.072  \\
    2      &        24.42  $\pm$         1.44   &       51.32  $\pm$         2.98   &        2.52  $\pm$         0.38   &        0.02  $\pm$         0.35   &       52.62  $\pm$        57.48   &       0.876  $\pm$        0.210  \\
    3      &        10.37  $\pm$        14.34   &       39.41  $\pm$         6.39   &       15.86  $\pm$         6.07   &        0.04  $\pm$         0.40   &        5.09  $\pm$        36.28   &       0.112  $\pm$        0.105  \\
    4      &        38.52  $\pm$         7.56   &       27.24  $\pm$        13.86   &        2.52  $\pm$         0.27   &        0.02  $\pm$         0.32   &       84.62  $\pm$        86.51   &       0.333  $\pm$        0.373  \\
    5      &        12.94  $\pm$        11.25   &       17.34  $\pm$        11.55   &       14.63  $\pm$         5.96   &        0.41  $\pm$         0.16   &       29.04  $\pm$        21.81   &       0.003  $\pm$        0.054  \\
    6      &        29.39  $\pm$         0.74   &       12.51  $\pm$        14.79   &       11.52  $\pm$         4.56   &        0.72  $\pm$         0.08   &      -48.23  $\pm$        26.84   &       0.155  $\pm$        0.086  \\\\
    1      &        25.00  $\pm$         0.01   &       52.00  $\pm$         0.01   &        3.00  $\pm$         0.01   &        0.60  $\pm$         0.00   &      -70.01  $\pm$         0.56   &       1.000  $\pm$        0.003  \\
    2      &        27.00  $\pm$         0.02   &       46.00  $\pm$         0.02   &        3.20  $\pm$         0.02   &        0.65  $\pm$         0.01   &      -49.98  $\pm$         0.89   &       0.500  $\pm$        0.003  \\
    3      &        32.97  $\pm$         0.12   &       38.01  $\pm$         0.08   &        3.99  $\pm$         0.09   &        0.70  $\pm$         0.02   &      -10.07  $\pm$         3.30   &       0.099  $\pm$        0.003  \\
    4      &        35.01  $\pm$         6.15   &       36.91  $\pm$         7.17   &        3.81  $\pm$         2.00   &        0.59  $\pm$         0.45   &       -1.59  $\pm$        93.35   &       0.001  $\pm$        0.002  \\
    5      &        35.86  $\pm$         4.82   &       36.75  $\pm$         4.40   &        3.57  $\pm$         1.52   &        0.59  $\pm$         0.42   &       -3.48  $\pm$        86.50   &       0.001  $\pm$        0.002  \\
    6      &        36.00  $\pm$         0.03   &       27.99  $\pm$         0.04   &        5.00  $\pm$         0.04   &        0.75  $\pm$         0.01   &       69.93  $\pm$         0.73   &       0.300  $\pm$        0.002  \\
    7      &        22.97  $\pm$         0.28   &       19.99  $\pm$         0.17   &        7.96  $\pm$         0.24   &        0.80  $\pm$         0.02   &       19.99  $\pm$         2.77   &       0.050  $\pm$        0.001  \\\\

\tableline

\tableline
\end{tabular}
\tablecomments{The number of the sources was varied from three to seven. The uncertainties in each parameter correspond to the $3\sigma$ level.}
\end{center}
\end{table*}



\begin{table*}
\footnotesize
\begin{center}
  \caption{Model Fitting Parameters of the Gaussian Sources in the Case of the  Synthetic Image of J1 with Addition of Realistic Noise Pattern. \label{tab_J1_modfit_noise}}
  \begin{tabular}{@{}ccccccc@{}}
\tableline\tableline
   & $x$ & $y$ & $a$ & $\epsilon$ & $\theta$ & $I_0$ \\
& (pixel) & (pixel) & (pixel) & & (deg) & (Jy/beam) \\
\tableline
    1      &        11.00  $\pm$         0.01   &       71.04  $\pm$         0.03   &        5.44  $\pm$         0.03   &        0.90  $\pm$         0.00   &       74.85  $\pm$         0.24   &       2.005  $\pm$        0.009  \\
    2      &        15.60  $\pm$         0.05   &       49.57  $\pm$         0.10   &        9.21  $\pm$         0.11   &        0.85  $\pm$         0.01   &      -86.52  $\pm$         0.82   &       0.365  $\pm$        0.004  \\\\
    1      &        11.00  $\pm$         0.00   &       71.00  $\pm$         0.01   &        5.50  $\pm$         0.01   &        0.90  $\pm$         0.00   &       75.00  $\pm$         0.08   &       2.000  $\pm$        0.003  \\
    2      &        17.00  $\pm$         0.04   &       54.02  $\pm$         0.08   &        6.48  $\pm$         0.06   &        0.80  $\pm$         0.01   &       82.64  $\pm$         0.74   &       0.300  $\pm$        0.004  \\
    3      &        13.02  $\pm$         0.10   &       43.97  $\pm$         0.12   &        8.02  $\pm$         0.08   &        0.70  $\pm$         0.01   &       40.89  $\pm$         1.35   &       0.200  $\pm$        0.003  \\\\
    1      &        11.00  $\pm$         5.85   &       71.00  $\pm$         3.17   &        5.50  $\pm$         3.22   &        0.90  $\pm$         0.41   &       75.00  $\pm$        44.09   &       1.382  $\pm$        0.005  \\
    2      &        11.00  $\pm$         3.52   &       71.00  $\pm$         6.22   &        5.50  $\pm$         2.79   &        0.90  $\pm$         0.15   &       75.00  $\pm$        30.45   &       0.110  $\pm$        0.005  \\
    3      &        17.01  $\pm$         0.06   &       54.06  $\pm$         0.13   &        6.48  $\pm$         0.10   &        0.80  $\pm$         0.01   &       82.91  $\pm$         1.16   &       0.298  $\pm$        0.007  \\
    4      &        13.06  $\pm$         0.15   &       44.02  $\pm$         0.18   &        8.04  $\pm$         0.13   &        0.70  $\pm$         0.01   &       41.63  $\pm$         2.08   &       0.201  $\pm$        0.005  \\\\
    1      &        11.00  $\pm$        14.89   &       71.02  $\pm$        12.04   &        5.56  $\pm$         3.08   &        0.90  $\pm$         0.09   &       74.94  $\pm$        15.47   &       1.153  $\pm$        0.056  \\
    2      &        11.04  $\pm$        15.42   &       71.01  $\pm$        13.52   &        5.33  $\pm$        21.83   &        0.89  $\pm$         0.09   &       75.14  $\pm$        96.36   &       0.017  $\pm$        0.063  \\
    3      &        11.01  $\pm$        11.51   &       70.93  $\pm$        12.11   &        5.28  $\pm$         3.08   &        0.88  $\pm$         0.53   &       75.29  $\pm$       105.27   &       0.022  $\pm$        0.036  \\
    4      &        13.10  $\pm$         0.12   &       65.60  $\pm$         0.25   &        5.74  $\pm$         0.19   &        0.89  $\pm$         0.02   &       76.01  $\pm$         2.27   &       0.305  $\pm$        0.090  \\
    5      &        14.84  $\pm$         0.28   &       47.17  $\pm$         0.35   &        8.55  $\pm$         0.31   &        0.81  $\pm$         0.03   &       63.56  $\pm$         4.44   &       0.273  $\pm$        0.010  \\\\
    1      &        19.62  $\pm$        11.54   &       74.11  $\pm$        41.60   &        3.46  $\pm$        43.82   &        0.91  $\pm$         0.54   &       73.29  $\pm$        89.52   &       0.002  $\pm$        0.010  \\
    2      &        11.00  $\pm$        20.64   &       71.01  $\pm$        26.13   &        5.49  $\pm$        34.44   &        0.90  $\pm$         0.17   &       74.95  $\pm$        38.78   &       0.143  $\pm$        0.010  \\
    3      &        11.00  $\pm$         2.63   &       71.00  $\pm$         3.50   &        5.50  $\pm$         1.26   &        0.90  $\pm$         0.42   &       75.00  $\pm$        83.42   &       1.933  $\pm$        0.011  \\
    4      &        21.01  $\pm$        11.65   &       60.90  $\pm$        40.89   &        4.81  $\pm$        46.18   &        0.90  $\pm$         0.57   &       75.55  $\pm$        91.04   &       0.001  $\pm$        0.014  \\
    5      &        17.00  $\pm$         0.16   &       54.01  $\pm$         0.39   &        6.47  $\pm$         0.21   &        0.80  $\pm$         0.02   &       -2.09  $\pm$         1.73   &       0.297  $\pm$        0.017  \\
    6      &        13.05  $\pm$         0.42   &       44.04  $\pm$         0.60   &        8.02  $\pm$         0.29   &        0.70  $\pm$         0.03   &       41.73  $\pm$         5.26   &       0.202  $\pm$        0.013  \\\\
    1      &        23.99  $\pm$         4.21   &       82.17  $\pm$         5.84   &       11.71  $\pm$         2.36   &        0.90  $\pm$         0.34   &       77.29  $\pm$        47.42   &       0.001  $\pm$        0.013  \\
    2      &        11.00  $\pm$         1.53   &       71.01  $\pm$         2.94   &        5.50  $\pm$         2.31   &        0.90  $\pm$         0.13   &       74.97  $\pm$        20.25   &       0.097  $\pm$        0.010  \\
    3      &        11.00  $\pm$         3.67   &       71.00  $\pm$         9.60   &        5.49  $\pm$         2.05   &        0.90  $\pm$         0.53   &       75.02  $\pm$        80.05   &       0.961  $\pm$        0.017  \\
    4      &        20.03  $\pm$         5.86   &       60.09  $\pm$        15.96   &        8.91  $\pm$         3.75   &        0.59  $\pm$         0.73   &       17.44  $\pm$       123.92   &       0.001  $\pm$        0.015  \\
    5      &        17.03  $\pm$         4.84   &       54.14  $\pm$         9.93   &        6.55  $\pm$         3.51   &        0.80  $\pm$         0.49   &       82.69  $\pm$        67.82   &       0.003  $\pm$        0.018  \\
    6      &        16.94  $\pm$         1.77   &       53.87  $\pm$         4.43   &        6.32  $\pm$         1.63   &        0.79  $\pm$         0.29   &      -77.18  $\pm$        25.46   &       0.209  $\pm$        0.019  \\
    7      &        13.09  $\pm$         0.42   &       44.18  $\pm$         0.48   &        7.97  $\pm$         0.35   &        0.70  $\pm$         0.04   &       43.44  $\pm$         6.16   &       0.208  $\pm$        0.013  \\\\
\tableline
\end{tabular}
\tablecomments{The number of the sources was varied from two to seven. The uncertainties in each parameter correspond to the $3\sigma$ level.}
\end{center}
\end{table*}



\begin{table*}
\footnotesize
\begin{center}
  \caption{Model Fitting Parameters of the Gaussian Sources in the Case of the  Synthetic Image of J2 with Addition of Realistic Noise Pattern. \label{tab_J2_modfit_noise}}
  \begin{tabular}{@{}ccccccc@{}}
\tableline\tableline
   & $x$ & $y$ & $a$ & $\epsilon$ & $\theta$ & $I_0$ \\
& (pixel) & (pixel) & (pixel) & & (deg) & (Jy/beam) \\
\tableline
    1      &        25.20  $\pm$         0.05   &       52.55  $\pm$         0.09   &        2.80  $\pm$         0.14   &        0.59  $\pm$         0.07   &      -70.28  $\pm$         6.37   &       0.740  $\pm$        0.084  \\
    2      &        26.20  $\pm$         0.12   &       47.59  $\pm$         0.47   &        4.62  $\pm$         0.12   &        0.85  $\pm$         0.01   &      -59.15  $\pm$         2.57   &       0.597  $\pm$        0.047  \\
    3      &        35.72  $\pm$         0.04   &       28.35  $\pm$         0.06   &        5.93  $\pm$         0.06   &        0.77  $\pm$         0.01   &      -90.00  $\pm$         0.27   &       0.277  $\pm$        0.003  \\\\
    1      &        25.19  $\pm$         0.09   &       52.56  $\pm$         0.20   &        2.74  $\pm$         0.18   &        0.56  $\pm$         0.05   &      -74.65  $\pm$         5.19   &       0.685  $\pm$        0.123  \\
    2      &        26.12  $\pm$         0.33   &       47.90  $\pm$         0.83   &        4.72  $\pm$         0.41   &        0.85  $\pm$         0.05   &      -61.52  $\pm$         2.16   &       0.620  $\pm$        0.070  \\
    3      &        35.93  $\pm$         0.52   &       29.46  $\pm$         1.11   &        6.33  $\pm$         0.29   &        0.85  $\pm$         0.01   &      -89.90  $\pm$         2.84   &       0.234  $\pm$        0.014  \\
    4      &        30.35  $\pm$         1.19   &       23.72  $\pm$         0.71   &       10.36  $\pm$         0.54   &        0.95  $\pm$         0.01   &       28.21  $\pm$         2.95   &       0.077  $\pm$        0.018  \\\\
    1      &        24.99  $\pm$         0.01   &       52.08  $\pm$         0.06   &        2.97  $\pm$         0.02   &        0.59  $\pm$         0.01   &      -69.26  $\pm$         1.02   &       0.981  $\pm$        0.016  \\
    2      &        26.92  $\pm$         0.06   &       46.14  $\pm$         0.11   &        3.32  $\pm$         0.10   &        0.68  $\pm$         0.02   &      -52.07  $\pm$         1.24   &       0.514  $\pm$        0.010  \\
    3      &        33.09  $\pm$         0.11   &       37.92  $\pm$         0.09   &        3.90  $\pm$         0.11   &        0.69  $\pm$         0.03   &       -6.34  $\pm$         5.58   &       0.100  $\pm$        0.002  \\
    4      &        36.00  $\pm$         0.03   &       28.03  $\pm$         0.07   &        5.01  $\pm$         0.04   &        0.75  $\pm$         0.01   &       70.07  $\pm$         0.91   &       0.300  $\pm$        0.002  \\
    5      &        22.95  $\pm$         0.36   &       19.98  $\pm$         0.20   &        8.00  $\pm$         0.34   &        0.80  $\pm$         0.02   &       18.85  $\pm$         2.67   &       0.050  $\pm$        0.002  \\\\
    1      &        22.03  $\pm$         4.07   &       57.14  $\pm$         9.16   &        2.85  $\pm$         0.44   &        0.86  $\pm$         0.14   &       28.63  $\pm$        69.10   &       0.039  $\pm$        0.412  \\
    2      &        25.53  $\pm$         7.54   &       50.15  $\pm$        10.96   &        2.67  $\pm$         1.84   &        0.06  $\pm$         0.61   &      -47.56  $\pm$        22.53   &       1.120  $\pm$        0.852  \\
    3      &        26.01  $\pm$         7.48   &       43.48  $\pm$        15.16   &       17.36  $\pm$        11.51   &        0.04  $\pm$         0.40   &      -53.53  $\pm$        30.24   &       0.015  $\pm$        0.154  \\
    4      &        26.16  $\pm$         1.00   &       38.76  $\pm$        11.33   &       29.73  $\pm$        22.66   &        0.71  $\pm$         0.11   &      -88.43  $\pm$        16.76   &       0.071  $\pm$        0.756  \\
    5      &        31.65  $\pm$         4.64   &       37.26  $\pm$         6.21   &       20.23  $\pm$        14.37   &        0.89  $\pm$         0.63   &       77.80  $\pm$       130.15   &       0.014  $\pm$        0.146  \\
    6      &        32.13  $\pm$         6.73   &       12.29  $\pm$         6.60   &        2.95  $\pm$         4.79   &        0.72  $\pm$         0.10   &      -16.67  $\pm$        26.55   &       0.403  $\pm$        0.296  \\\\
    1      &        45.84  $\pm$         2.79   &       55.77  $\pm$         2.78   &       28.72  $\pm$         2.31   &        0.78  $\pm$         0.25   &      -43.40  $\pm$        58.53   &       0.000  $\pm$        0.000  \\
    2      &        41.63  $\pm$        10.18   &       53.99  $\pm$         5.73   &       10.71  $\pm$        23.06   &        0.77  $\pm$         0.28   &      -22.08  $\pm$        73.25   &       0.000  $\pm$        0.001  \\
    3      &        25.00  $\pm$         0.01   &       52.00  $\pm$         0.01   &        3.00  $\pm$         0.01   &        0.60  $\pm$         0.00   &      -70.09  $\pm$         0.63   &       0.999  $\pm$        0.003  \\
    4      &        27.00  $\pm$         0.02   &       46.00  $\pm$         0.02   &        3.20  $\pm$         0.02   &        0.65  $\pm$         0.01   &      -50.32  $\pm$         0.99   &       0.500  $\pm$        0.003  \\
    5      &        33.03  $\pm$         0.12   &       37.97  $\pm$         0.09   &        3.99  $\pm$         0.10   &        0.70  $\pm$         0.02   &      -10.92  $\pm$         3.53   &       0.101  $\pm$        0.002  \\
    6      &        35.99  $\pm$         0.03   &       27.99  $\pm$         0.05   &        4.99  $\pm$         0.04   &        0.75  $\pm$         0.01   &       69.81  $\pm$         0.82   &       0.301  $\pm$        0.002  \\
    7      &        22.85  $\pm$         0.30   &       19.93  $\pm$         0.18   &        7.87  $\pm$         0.26   &        0.79  $\pm$         0.02   &       18.48  $\pm$         2.86   &       0.050  $\pm$        0.002  \\\\
\tableline
\end{tabular}
\tablecomments{The number of the sources was varied from two to seven. The uncertainties in each parameter correspond to the $3\sigma$ level.}
\end{center}
\end{table*}



\begin{table*}
\footnotesize
\begin{center}
  \caption{Comparison between the Relative Errors in the Parameters of the Gaussian Sources of the Noisy Jet J1 Obtained from our CE Technique and the AIPS Task IMFIT (Values in Parentheses). \label{J1_CE_IMFIT}}
  \begin{tabular}{@{}ccccccc@{}}
\tableline\tableline
   & $\epsilon_x$ & $\epsilon_y$ & $\epsilon_a$ & $\epsilon_\epsilon$ & $\epsilon_\theta$ & $\epsilon_{I_0}$ \\
& (\%) & (\%) & (\%) & (\%) & (\%) & (\%) \\
\tableline
    1      &      0.000  $\pm$      0.000   &     0.000  $\pm$      0.000   &     0.000  $\pm$      0.000   &     0.000  $\pm$      0.000   &     0.000  $\pm$      0.053   &     0.000  $\pm$      0.050 \\
     & (0.005  $\pm$      0.013)   & (0.000  $\pm$      0.013)   & (0.012  $\pm$      0.203)   &(0.001  $\pm$      0.057)   &(0.004  $\pm$      0.100)   &(0.000  $\pm$      0.172) \\\\
    2      &      0.000  $\pm$      0.118   &     0.037  $\pm$      0.056   &     0.308  $\pm$      0.462   &     0.000  $\pm$      0.000   &     0.470  $\pm$      0.410   &     0.000  $\pm$      0.667 \\
     & (0.035  $\pm$      0.268)   & (0.050  $\pm$      0.137)   & (0.303  $\pm$      1.351)   &(0.290  $\pm$      0.928)   &(0.357  $\pm$      1.065)   &(0.120  $\pm$      1.150) \\\\
    3      &      0.154  $\pm$      0.308   &     0.068  $\pm$      0.091   &     0.250  $\pm$      0.375   &     0.000  $\pm$      0.000   &     0.341  $\pm$      1.415   &     0.000  $\pm$      0.500 \\
     & (0.177  $\pm$      0.942)   & (0.036  $\pm$      0.266)   & (0.209  $\pm$      2.031)   &(0.182  $\pm$      2.520)   &(0.012  $\pm$      4.929)   &(0.165  $\pm$      1.725) \\\\

\tableline
\end{tabular}
\tablecomments{The uncertainties in each parameter correspond to the $3\sigma$ level.}
\end{center}
\end{table*}



\begin{table*}
\scriptsize
\begin{center}
  \caption{Comparison between the Relative Errors in the Parameters of the Gaussian Sources of the Noisy Jet J3 Obtained from our CE Technique and the AIPS Task IMFIT (Values in Parentheses). \label{J3_CE_IMFIT}}
  \begin{tabular}{@{}ccccccc@{}}
\tableline\tableline
   & $\epsilon_x$ & $\epsilon_y$ & $\epsilon_a$ & $\epsilon_\epsilon$ & $\epsilon_\theta$ & $\epsilon_{I_0}$ \\
& (\%) & (\%) & (\%) & (\%) & (\%) & (\%) \\
\tableline
    1      &      0.000  $\pm$      0.013   &     0.058  $\pm$      0.013   &     0.333  $\pm$      0.111   &     0.000  $\pm$      0.000   &     0.543  $\pm$      0.238   &     0.700  $\pm$      0.167 \\
     & (4.000  $\pm$      0.009)   & (1.941  $\pm$      0.009)   & (0.135  $\pm$      0.187)   &(0.194  $\pm$      0.398)   &(0.354  $\pm$      0.407)   &(0.180  $\pm$      0.159) \\\\
    2      &      0.148  $\pm$      0.037   &     0.130  $\pm$      0.029   &     1.250  $\pm$      0.312   &     1.538  $\pm$      0.513   &     1.420  $\pm$      0.413   &     1.200  $\pm$      0.267 \\
     & (4.030  $\pm$      0.064)   & (1.735  $\pm$      0.039)   & (2.514  $\pm$      0.727)   &(5.344  $\pm$      1.392)   &(0.604  $\pm$      2.138)   &(49.778  $\pm$      0.318) \\\\
    3      &      0.000  $\pm$      0.019   &     0.071  $\pm$      0.036   &     0.000  $\pm$      0.200   &     0.000  $\pm$      0.000   &     0.100  $\pm$      0.248   &     0.000  $\pm$      0.222 \\
     & (3.669  $\pm$      0.059)   & (4.843  $\pm$      0.098)   & (8.331  $\pm$      0.685)   &(2.225  $\pm$      0.545)   &(8.801  $\pm$      0.679)   &(1.293  $\pm$      0.530) \\\\
    4      &      0.043  $\pm$      0.333   &     0.050  $\pm$      0.200   &     0.500  $\pm$      0.875   &     1.250  $\pm$      0.417   &     6.800  $\pm$      2.867   &     0.000  $\pm$      0.667 \\
     & (13.561  $\pm$      0.077)   & (123.795  $\pm$      0.093)   & (59.578  $\pm$      0.304)   &(16.249  $\pm$      0.927)   &(350.870  $\pm$      4.270)   &(398.780  $\pm$      3.180) \\\\

\tableline
\end{tabular}
\tablecomments{The uncertainties in each parameter correspond to the $1\sigma$ level.}
\end{center}
\end{table*}



\begin{table*}
\footnotesize
\begin{center}
  \caption{Model Fitting Parameters of the Elliptical Gaussian Sources in the Case of 15 GHz VLBA Image of the BL Lac OJ 287 Obtained in 1996 May 27. \label{tab_OJ287_modfit_ellipt}}
  \begin{tabular}{@{}ccccccc@{}}
\tableline\tableline
   & $x$ & $y$ & $a$ & $\epsilon$ & $\theta$ & $I_0$\\
& (pixel) & (pixel) & (pixel) & & (deg) & (Jy/beam) \\
\tableline
    1      &       253.65  $\pm$         0.02   &      257.75  $\pm$         0.01   &        4.94  $\pm$         0.01   &        0.85  $\pm$         0.00   &      -66.65  $\pm$         0.21   &       1.023  $\pm$        0.021  \\
    2      &       257.79  $\pm$         0.26   &      257.39  $\pm$         0.03   &        5.35  $\pm$         0.03   &        0.65  $\pm$         0.01   &      -43.55  $\pm$         0.93   &       0.251  $\pm$        0.018  \\\\
    1      &       253.46  $\pm$         0.03   &      257.81  $\pm$         0.03   &        4.92  $\pm$         0.01   &        0.85  $\pm$         0.00   &      -67.30  $\pm$         0.09   &       0.963  $\pm$        0.032  \\
    2      &       257.08  $\pm$         0.39   &      257.34  $\pm$         0.04   &        5.10  $\pm$         0.06   &        0.69  $\pm$         0.00   &      -56.31  $\pm$         1.91   &       0.330  $\pm$        0.022  \\
    3      &       267.21  $\pm$         1.71   &      257.19  $\pm$         0.45   &        4.97  $\pm$         0.18   &        0.67  $\pm$         0.33   &      -66.06  $\pm$        19.54   &       0.026  $\pm$        0.005  \\\\
    1      &       253.47  $\pm$         0.02   &      257.81  $\pm$         0.02   &        4.92  $\pm$         0.01   &        0.85  $\pm$         0.00   &      -67.26  $\pm$         0.08   &       0.979  $\pm$        0.016  \\
    2      &       257.26  $\pm$         0.19   &      257.33  $\pm$         0.02   &        5.10  $\pm$         0.03   &        0.69  $\pm$         0.00   &      -56.36  $\pm$         1.09   &       0.319  $\pm$        0.011  \\
    3      &       267.64  $\pm$         0.62   &      257.22  $\pm$         0.37   &        4.94  $\pm$         0.24   &        0.74  $\pm$         0.14   &      -69.89  $\pm$         8.21   &       0.024  $\pm$        0.002  \\
    4      &       286.83  $\pm$        27.10   &      247.32  $\pm$        18.43   &        8.24  $\pm$         3.91   &        0.34  $\pm$         0.47   &      -26.31  $\pm$       101.36   &       0.001  $\pm$        0.001  \\\\
    1      &       253.48  $\pm$         0.04   &      257.79  $\pm$         0.06   &        4.92  $\pm$         0.01   &        0.85  $\pm$         0.00   &      -67.19  $\pm$         0.22   &       0.993  $\pm$        0.048  \\
    2      &       257.41  $\pm$         0.52   &      257.36  $\pm$         0.08   &        5.11  $\pm$         0.04   &        0.70  $\pm$         0.03   &      -56.06  $\pm$         1.65   &       0.309  $\pm$        0.030  \\
    3      &       267.80  $\pm$         0.49   &      257.22  $\pm$         0.65   &        4.94  $\pm$         0.41   &        0.78  $\pm$         0.13   &      -69.20  $\pm$         7.94   &       0.023  $\pm$        0.003  \\
    4      &       270.60  $\pm$        13.21   &      261.52  $\pm$         8.71   &        6.22  $\pm$         4.06   &        0.60  $\pm$         0.71   &       46.32  $\pm$       103.67   &       0.001  $\pm$        0.002  \\
    5      &       286.76  $\pm$        46.74   &      251.15  $\pm$        16.89   &       11.51  $\pm$        16.73   &        0.63  $\pm$         0.78   &        4.61  $\pm$        95.79   &       0.001  $\pm$        0.002  \\\\
    1      &       249.58  $\pm$        35.10   &      262.14  $\pm$         5.32   &       10.46  $\pm$        12.18   &        0.40  $\pm$         0.27   &      -49.96  $\pm$       171.24   &       0.209  $\pm$        0.459  \\
    2      &       250.44  $\pm$         6.69   &      240.54  $\pm$        38.20   &        4.49  $\pm$         0.96   &        0.70  $\pm$         0.34   &      -68.66  $\pm$         3.07   &       0.343  $\pm$        1.392  \\
    3      &       256.97  $\pm$        23.46   &      240.91  $\pm$        35.56   &        4.28  $\pm$         1.16   &        0.66  $\pm$         0.20   &      -32.52  $\pm$        79.82   &       0.121  $\pm$        0.218  \\
    4      &       258.72  $\pm$        18.19   &      257.04  $\pm$         5.17   &        4.52  $\pm$         2.42   &        0.44  $\pm$         0.29   &       -7.00  $\pm$        63.81   &       0.469  $\pm$        1.033  \\
    5      &       265.66  $\pm$        37.73   &      261.34  $\pm$        27.97   &        6.89  $\pm$         2.03   &        0.74  $\pm$         0.90   &       43.50  $\pm$       140.40   &       0.001  $\pm$        0.001  \\
    6      &       272.93  $\pm$        34.85   &      255.07  $\pm$         5.02   &       13.00  $\pm$        17.47   &        0.39  $\pm$         0.66   &      -12.24  $\pm$        97.00   &       0.114  $\pm$        0.461  \\\\
    1      &       246.40  $\pm$         1.61   &      262.42  $\pm$         1.05   &        6.48  $\pm$         0.35   &        0.42  $\pm$         0.10   &       12.27  $\pm$        17.84   &       0.428  $\pm$        0.146  \\
    2      &       252.84  $\pm$         2.89   &      231.15  $\pm$         6.23   &       11.85  $\pm$         1.37   &        0.80  $\pm$         0.07   &      -23.07  $\pm$        14.37   &       0.151  $\pm$        0.033  \\
    3      &       256.26  $\pm$         2.03   &      257.95  $\pm$         0.68   &        4.01  $\pm$         0.58   &        0.21  $\pm$         0.10   &       17.13  $\pm$         9.89   &       0.685  $\pm$        0.153  \\
    4      &       264.70  $\pm$         1.48   &      263.77  $\pm$         1.50   &       13.26  $\pm$         1.63   &        0.00  $\pm$         0.14   &       75.57  $\pm$        14.68   &       0.001  $\pm$        0.001  \\
    5      &       268.39  $\pm$         2.66   &      271.38  $\pm$         3.60   &        4.02  $\pm$         1.17   &        0.27  $\pm$         0.09   &       -0.70  $\pm$         9.55   &       0.001  $\pm$        0.000  \\
    6      &       300.81  $\pm$         9.59   &      261.40  $\pm$         0.90   &       19.62  $\pm$         3.29   &        0.83  $\pm$         0.02   &      -17.38  $\pm$        10.27   &       0.156  $\pm$        0.023  \\
    7      &       303.79  $\pm$         8.29   &      234.27  $\pm$         5.07   &       29.86  $\pm$         5.68   &        0.10  $\pm$         0.12   &       69.01  $\pm$        29.72   &       0.001  $\pm$        0.005  \\\\

\tableline
\end{tabular}
\tablecomments{The number of sources was varied from two to seven. Component 1 refers to the compact core of OJ\,287. The uncertainties in each parameter correspond to the $3\sigma$ level.}
\end{center}
\end{table*}



\begin{table*}
\footnotesize
\begin{center}
  \caption{Comparison between the Values of the Parameters Found in This Work for $N_\mathrm{s}=3$ and Those Published in \citet{lis09b} (in Parentheses) for the 15 GHz Interferometric Map of OJ\,287 obtained in 1996 May 27. \label{table_comparative_OJ287_960527_v2}}
  \begin{tabular}{@{}ccccc@{}}
\tableline\tableline
   & $F$ & $r$ & $\eta$ & $a_\mathrm{FWHM}$  \\
& (Jy) & (mas) & (deg) & (\%) \\
\tableline
    1      &       0.96  $\pm$       0.03   &     -   &      -   &      0.06  $\pm$       0.00 \\
     & (1.11  $\pm$       0.06)   & -   & -   &(0.08  $\pm$       0.01) \\\\
    2      &       0.49  $\pm$       0.03   &      0.37  $\pm$       0.05   &    -97.40  $\pm$       1.45   &      0.59  $\pm$       0.01 \\
     & (0.31  $\pm$       0.02)   & (0.47  $\pm$       0.05)   & (-95.50  $\pm$       9.55)   &(0.33  $\pm$       0.03) \\\\
    3      &       0.04  $\pm$       0.02   &      1.38  $\pm$       0.24   &    -92.60  $\pm$       2.66   &      0.58  $\pm$       0.19 \\
     & (0.07  $\pm$       0.00)   & (1.10  $\pm$       0.05)   & (-92.20  $\pm$       9.22)   &(0.53  $\pm$       0.05) \\\\

\tableline
\end{tabular}
\tablecomments{ The uncertainties concerning CE results correspond to 1$\sigma$-value, which was obtained from standard propagation of the errors.}
\end{center}
\end{table*}


\end{document}